\documentclass[twoside]{article}

\usepackage{qic2}

\usepackage{graphicx}
\usepackage{bm}
\usepackage{braket}
\usepackage{comment}
\usepackage{amsmath}
\usepackage{amssymb}
\usepackage{subfigure}
\usepackage{cite}

\newcommand{\myeqnref}[1]{Eq.~\eqref{#1}}
\newcommand{\field}[1]{\mathbb{#1}} 

\newtheorem{claim}{Claim}

\begin{document}
\setlength{\textheight}{8.0truein}    

\runninghead{Security proof of quantum key distribution with detection efficiency mismatch}
            {C.-H. F. Fung, K. Tamaki, B. Qi, H.-K. Lo, and X. Ma}

\normalsize\textlineskip
\thispagestyle{empty}
\setcounter{page}{1}

\vspace*{0.88truein}

\alphfootnote

\fpage{1}

\centerline{\bf
SECURITY PROOF OF QUANTUM KEY DISTRIBUTION
}
\vspace*{0.035truein}
\centerline{\bf WITH DETECTION EFFICIENCY MISMATCH}
\vspace*{0.37truein}
\centerline{\footnotesize
CHI-HANG FRED FUNG$^1$, KIYOSHI TAMAKI$^2$}
\vspace*{0.015truein}
\centerline{\footnotesize 
BING QI$^3$, HOI-KWONG LO$^3$, and XIONGFENG MA$^4$}
\vspace*{10pt}
\centerline{\footnotesize\it $^1$Department of Physics and Center of Computational and Theoretical Physics,}
\baselineskip=10pt
\centerline{\footnotesize\it  University of Hong Kong, Pokfulam Road, Hong Kong, China}
\vspace*{10pt}
\centerline{\footnotesize\it $^2$NTT Basic Research Laboratories, NTT Corporation,}
\baselineskip=10pt
\centerline{\footnotesize\it 3-1, Morinosato Wakamiya Atsugi-Shi, Kanagawa, 243-0198, Japan;}
\baselineskip=10pt
\centerline{\footnotesize\it CREST, JST Agency, 4-1-8 Honcho, Kawaguchi, Saitama, 332-0012, Japan}
\vspace*{10pt}
\centerline{\footnotesize\it $^3$Center for Quantum Information and Quantum Control,}
\baselineskip=10pt
\centerline{\footnotesize\it Department of Physics and Department of Electrical \& Computer Engineering,}
\baselineskip=10pt
\centerline{\footnotesize\it University of Toronto, Toronto, Ontario, M5S 3G4, Canada}
\vspace*{10pt}
\centerline{\footnotesize\it $^4$Institute for Quantum Computing, University of Waterloo,}
\baselineskip=10pt
\centerline{\footnotesize\it 200 University
Ave. W., Waterloo, Ontario, N2L 3G1, Canada}
\vspace*{0.225truein}
\publisher{(received date)}{(revised date)}

\vspace*{0.21truein}

\abstracts{
In theory, quantum key distribution (QKD) offers unconditional
security based on the laws of physics.
However, as
demonstrated in recent quantum hacking theory
and experimental papers, detection efficiency loophole can be fatal
to the security of practical QKD systems.
Here, we describe the physical origin of detection efficiency mismatch 
in various domains including spatial, spectral, and time domains and in
various experimental set-ups.
More importantly, we prove the unconditional
security of QKD even with detection efficiency mismatch.
We explicitly show how the key generation rate is characterized by
the maximal detection efficiency ratio between the two detectors.
Furthermore, we prove that by randomly switching the bit assignments of the detectors,
the effect of detection efficiency mismatch can be completely eliminated. 
}{}{}

\vspace*{10pt}

\vspace*{3pt}

\vspace*{1pt}\textlineskip    
\section{Introduction}        

Quantum key distribution (QKD)  \cite{Bennett1984,Ekert1991,Gisin2002} provides a way for two legitimate users, Alice and Bob, to share a secret key that is secure against an eavesdropper, Eve, who is only restricted by quantum mechanics.
After the successful sharing of the secret key, Alice and Bob can then use it in cryptographic applications such as secure communications and authentication.
Many previous unconditional security proofs for QKD \cite{Mayers2001,Biham2000,Lo1999,Shor2000,Ben-Or2002,Koashi2003,Renner2005,Renner2005c,Kraus2005,Gottesman2003,Chau2002,Christandl2004}
consider the case with perfect devices, such as perfect single-photon sources.
However, realistic devices are never perfect.
For example, weak coherent sources are widely used in practice to simulate single-photon emissions.
Thus, security proofs have to be extended to cover these practical imperfections in order to guarantee the security of a practical system.
Recently,
the use of weak coherent sources and threshold detectors have been considered by various security proofs
\cite{Inamori2005,Gottesman2004,Koashi2005b,Koashi2006,Kraus2007,Hayashi2007,Lo200707}.
In this paper, we consider another realistic imperfection found in detectors, which is the dependency of detection efficiency
on some auxiliary dimension such as the arrival time of signal;
furthermore, the efficiencies of the two detectors can be different.
We prove the unconditional security for this case.
In this paper, we consider a QKD scheme where Bob uses two separate detectors for detecting bits ``0'' and ``1''.

The general physical problem we are facing in practical QKD system with two detectors is
the detection efficiency loophole \cite{Marshall1983,Ferrero1990}.
This loophole underlies not only
fundamental physics like Bell inequalities, but also applied
technology like QKD.
Also, in practice, it is hard to build
two detectors that have exactly the same characteristics 
(e.g. frequency, time and spatial responses).
Moreover,
the problem of
detection efficiency mismatch is important not only for gated detectors,
but also for detectors with dead-times.
Indeed, standard InGaAs detectors
(and even SSPDs) have dead-times.
That is to say that after a detection
event, a detector becomes inactive for some prescribed time duration.
If Bob
registers a signal during a time period when one detector is dead while the
other one is still active, Eve will be able to infer which detector clicks.
Our paper is an illustration of how one can proceed to handle
this general problem of detection efficiency mismatch in the security of QKD.

In fact, 
attacks drawing on detection efficiency mismatch have been proposed before: the faked states attack \cite{Makarov2006,Makarov2007} and the time-shift attack \cite{Qi2007a}.
The faked-states attack \cite{Makarov2006,Makarov2007} is an intercept-and-resend attack, which is
challenging to implement because of synchronization and interferometric stability
issues. For this reason, the faked-states attack has never been
implemented experimentally.
In contrast, the time-shift attack \cite{Qi2007a} does
not involve any measurement and is, therefore, easier to implement in this regard.
Indeed, we have successfully demonstrated the time-shift attack experimentally on a commercial QKD system~\cite{Zhao2007a}, 
which indicates that detection efficiency mismatch is a serious realistic issue that can have a fatal effect on practical 
security%
\footnote{
Note that to implement the time-shift attack successfully, one may need to implement time compression of signals.
The signals being compressed
here are strong classical pulses from Bob to Alice in a plug-and-play system. Fortunately,
compression of strong classical pulses is a mature standard technique
that has been used in industries for decades (see, e.g., \cite{Ghatak1998}).
For simplicity, the experiment in Ref.~\cite{Zhao2007a} did not implement this mature standard technique but simulated this compression process by replacing the original laser source with one that produces narrower pulses.
}.

The success of the faked-states attack and the time-shift attack relies on the existence of detection efficiency mismatch and the assumption that Alice and Bob distill keys using a standard security proof that ignores this efficiency mismatch.
As we have noted before \cite{Zhao2007a,Lo2007Tropical}, 
once Alice and Bob are aware of an attack based on some imperfection, it may not be too difficult for them to devise counter measures against it.
On the other hand, counter-measures against eavesdropping can often
lead to new loopholes and/or be defeated by new eavesdropping
attacks. For instance, the four-state setting \cite{Nielsen2001,LaGasse2005} by Bob (intended to counter attacks based on detection efficiency mismatch) can be defeated
by a combination of the large-pulse attack \cite{Gisin2005} and the time-shift attack
\cite{Lo2007Tropical,Lydersen2008}.
Instead of implementing physical 
counter-measures, another way to guard against attacks based on detection efficiency mismatch is for Alice and Bob to take this imperfection into account when distilling keys.
Because of this imperfection, an extra amount of privacy amplification will be needed.
In this paper, 
we provide an unconditional security proof that takes detection efficiency mismatch into account, and by using this proof, Alice and Bob will be able to determine the right amount of privacy amplification needed to remove Eve's information on the final key even in the presence of 
detection efficiency mismatch.
We note that our proof is valid even if Eve performs the most general attack that correlates different  signal transmissions.

In this paper, we consider the BB84 protocol \cite{Bennett1984}, but the idea of our security proof can be used for other protocols.
For simplicity, for much of this paper, we consider that 
the input to Bob are single-photon signals.
The case of
multi-photon input signals will be discussed in Sec.~\ref{sec-multi-photon-input}.
For each transmission by Alice, Eve resends two systems to Bob: one system carrying the bit information and the other system representing the auxiliary system that the efficiencies of the detectors respond to.
We assume in this paper that
the information-carrying system is a qubit.
On the other hand, the auxiliary system
can have arbitrary dimension and is completely controlled by Eve in order to produce different effects on Bob due to imperfect detection efficiencies.
More specifically, Eve can send Bob an arbitrary state in this auxiliary system to induce different probabilities of detection for the two detectors.
Overall, Bob's system is represented by an enlarged quantum space and
the bit information only lives in a qubit subspace of it.
Thus, our security proof is one that applies to a protocol with an enlarged quantum space 
(which has recently been formalized in Ref.~\cite{Gelles2007}).
Also, we note that this auxiliary system 
essentially acts as a shield \cite{Horodecki2005} that protects Alice and Bob's key from Eve.
Thus, our proof serves as an example of shield analysis.

Bob's measurement operates on both the qubit system (which carries the information) and the auxiliary system (which affects the detection efficiencies) and we need a detector 
model that incorporates both systems.
In this paper, we do not attempt to treat the most general measurements but to assume a slightly more restrictive detector model that does not couple the two systems.
In other words, our model applies whenever the information-carrying dimension does not couple with the efficiency-affecting dimension in the detection process.
This detector model by no means covers all possible detection scenarios\footnote{For instance, our detector model does not cover the case when
Bob's Hilbert space is a qutrit.} \ 
but is general enough to include many interesting ones (such as those described in Sec.~\ref{sec-attacks}) that arise in practice.
It is important to note that even though the detector model does not couple the two systems, they can be arbitrarily coupled (entangled)
in the input state sent by Eve.

To get a glimpse at our detector model,
let's first consider the case
where there is no Eve and no efficiency mismatch.
In this case,  
the entanglement view of Alice's and Bob's systems is simply the perfect EPR pair $\ket{00}_{AB}+\ket{11}_{AB}$.
Now, supposing that there is efficiency mismatch and each detector $i$ has a constant efficiency $\eta_i$,
the state can then be represented as $\sqrt{\eta_0} \ket{00}_{AB} + \sqrt{\eta_1} \ket{11}_{AB}$, which can be regarded as a non-uniform EPR pair.
This simple case is a special case of what we consider in our security proof.
In fact, we consider a detector model that goes beyond the scalar efficiency model.
Our model can incorporate, for example, states of the form
$ \ket{00}_{AB} \otimes (F_0 \ket{\gamma}_T)  + \ket{11}_{AB} \otimes (F_1 \ket{\gamma}_T)$,
where $F_i$ is the filtering operation of detector $i$ acting on system $T$ modeling the detector's efficiency response and $\ket{\gamma}$ is an arbitrary state chosen by Eve to induce different effects on the two detectors.
The difficulty in proving the security is related to the 
non-uniformity in bits ``0'' and ``1'' of the overall state
and the fact the initial state in system $T$ is arbitrary.
Nevertheless, we propose a technique that proves the security even though Eve can choose any state in system $T$.

Our proof considers that Alice uses a single-photon source.
However, our result immediately applies to the case when Alice uses a phase-randomized weak coherent source, by using the results of Ref.~\cite{Gottesman2004,Koashi2005b}.
Our proof is founded on Koashi's general security proof 
based on the uncertainty principle~\cite{Koashi2005b}.
The essential strategy is to estimate how certain Bob can predict Alice's measurement outcomes in the basis conjugate to the key-generating basis.
To do this, we consider a virtual protocol in which Bob performs a virtual measurement on his enlarged quantum space where the information-carrying qubit is embedded.
His measurement result then gives a good prediction of Alice's.

Intuition tells us that the larger the mismatch between the two detectors is, the lower the key generation rate becomes.
We confirm this belief in our paper and explicitly quantify the exact effect of the mismatch on the key generation rate.
We show that the 
maximum efficiency ratio between the two detectors is directly related to 
the key generation rate [see Eqs.~\eqref{eqn-psucc-suboptimal1}, \eqref{eqn-ep-suboptimal1}, and \eqref{eqn-noisy-final-key-rate}].

There has been much interest in security proofs of QKD based on
fundamental principles such as no-signaling faster than the speed of light
and Bell's inequality violations \cite{Acin2006,Scarani2006}. In particular, the idea of device-independent
security proofs has been proposed. So far a complete proof of security
along this line is still missing and such a proof idea only applies to specialized
attacks such as collective attacks.
We remark that this paper, following our earlier time-shift attack
papers \cite{Qi2007a,Zhao2007a}, serves to highlight a fundamental weakness of
device-independent security proof. Even if such a proof can be constructed
in future for the most general attack, it will {\it not} apply to most
current practical QKD set-ups. This is because of the existence
of detection efficiency loophole and Eve's potential ability to
manipulate signals in the auxiliary domain to bias the
relative detection efficiency between two detectors.
(Notice, however, that, as we will show in Sec.~\ref{sec-fourphase},
a four-phase setting idea can equalize the detection efficiency
of two detectors.)

This paper is organized as follows.
In Sec.~\ref{sec-attacks}, we give an overview of the physical origin of detection efficiency mismatch and how Eve might exploit this mismatch.
We then introduce our detector model in Sec~\ref{sec-model}.
In Sec.~\ref{sec-noiseless}, we consider the security proof for the noiseless case, in which Eve does not introduce any bit or phase errors but only intervenes with the auxiliary dimension.
The study of the noiseless case is instructive as it illustrates clearly our proof technique that is shared across both the noiseless and noisy cases.
In Sec.~\ref{sec-noisy}, we provide the unconditional security proof for the noisy case, in which Eve introduces bit and phase errors, in addition to intervening with the auxiliary dimension.
In Sec.~\ref{sec-fourphase}, we show 
how randomly switching the bit assignments of the two detectors for each quantum signal
(similar to the scheme proposed by Refs.~\cite{Nielsen2001,LaGasse2005}) 
can completely eliminate the effect of detection efficiency mismatch.
In Sec.~\ref{sec-multi-photon}, we discuss how to handle multi-photon signals in our proof, with the help of the results of other papers.
Finally, we conclude in Sec.~\ref{sec-conclusion}.

\section{Practical detection efficiency mismatch and attacks\label{sec-attacks}}
\noindent
In this section, we consider detection efficiency mismatch in realistic QKD setups and possible attack strategies that are based on detection efficiency mismatch.
Without a security proof that takes into account of detection efficiency mismatch, these attacks may compromise the security of  QKD systems.
The security proof provided later in this paper takes detection efficiency mismatch into account and allows Alice and Bob to defeat these attacks.

\subsection{
Time-domain detection efficiency mismatch
}
\noindent
In a typical fiber-based QKD system operating at 1550nm wavelength, in order to minimize the effect of dark counts, the two InGaAs detectors are usually gated to be active in a narrow time window in which signals are expected to arrive.
However, due to the different responses of the two detectors and also the asymmetry between the two detection channels,
the detectors may have different efficiencies over time.
\begin{figure}
\includegraphics[width=1\columnwidth]{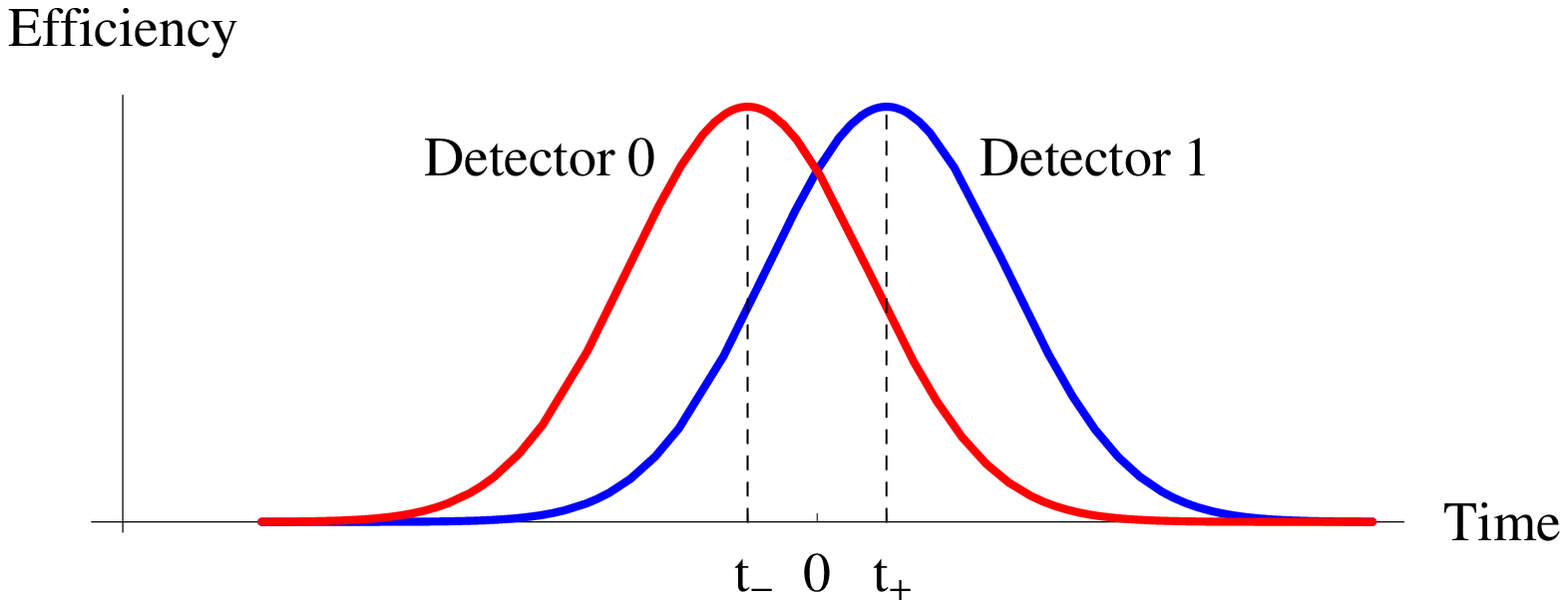}
\fcaption{\label{fig:mismatch-time-shift1}
Mismatch in efficiency between two detectors.
Due to the asymmetry between the two detection channels
the time responses of the two detectors may not be identical.
Under normal operation, signals are expected to arrive at time $0$.
On the other hand, Eve can launch a ``time-shift attack'' in which she shifts the arrival times of the signals to say times $t_{-}$ or $t_{+}$ in order to subject the signals to different probabilities of being detected as bit ``0'' or bit ``1''.
Thus, because of this difference in the detection probabilities of bit ``0'' and bit ``1'', Eve can learn some information about the key.
}
\end{figure}
Fig.~\ref{fig:mismatch-time-shift1} illustrates a mismatch in efficiency between two detectors.
Typically, the open windows of the two detectors are larger than the width of the laser pulses.
Thus, under normal operation, the signals arrive near the centre of the two open windows so that the efficiencies for detecting bit ``0'' and bit ``1'' are similar.
However, it is possible for Eve to time shift the input signals to Bob causing a mismatch in the efficiencies.
In Ref.~\cite{Qi2007a}, we proposed a ``time-shift attack'' that basically draws on this efficiency mismatch due to time shifting.
Essentially, Eve time shifts the signals entering Bob to say time $t_{-}$.
Whenever Bob announces that he has a detection, Eve knows that it is likely that detector 0 has clicked because it has a higher efficiency at time $t_{-}$.
Thus, Eve obtains some information about the bit value.
In the extreme case that at a particular time shift, the efficiency of one detector is positive and efficiency of the other detector is zero, Eve knows the bit value exactly because only one detector can ever produce a click.
Standard security proofs that ignore efficiency mismatch may allow Eve to steal information in this way.
On the other hand, the security proof provided in this paper takes this mismatch into account and applies enough privacy amplification depending on the mismatch to remove Eve's information.
We remark that a successful experimental demonstration of this attack has been performed by us \cite{Zhao2007a}.

Note that when multiple pulses arrive in the same detection window, the efficiency of the detection system may depend on the relative phases of these multiple pulses.
This may happen when there exist multiple reflection sites in the detection channel.
We will establish a detector model in Sec.~\ref{sec-model} that is general enough to incorporate this correlation.

\subsection{
Space-domain detection efficiency mismatch
}
\begin{figure}
\begin{center}
\includegraphics[width=.5\columnwidth]{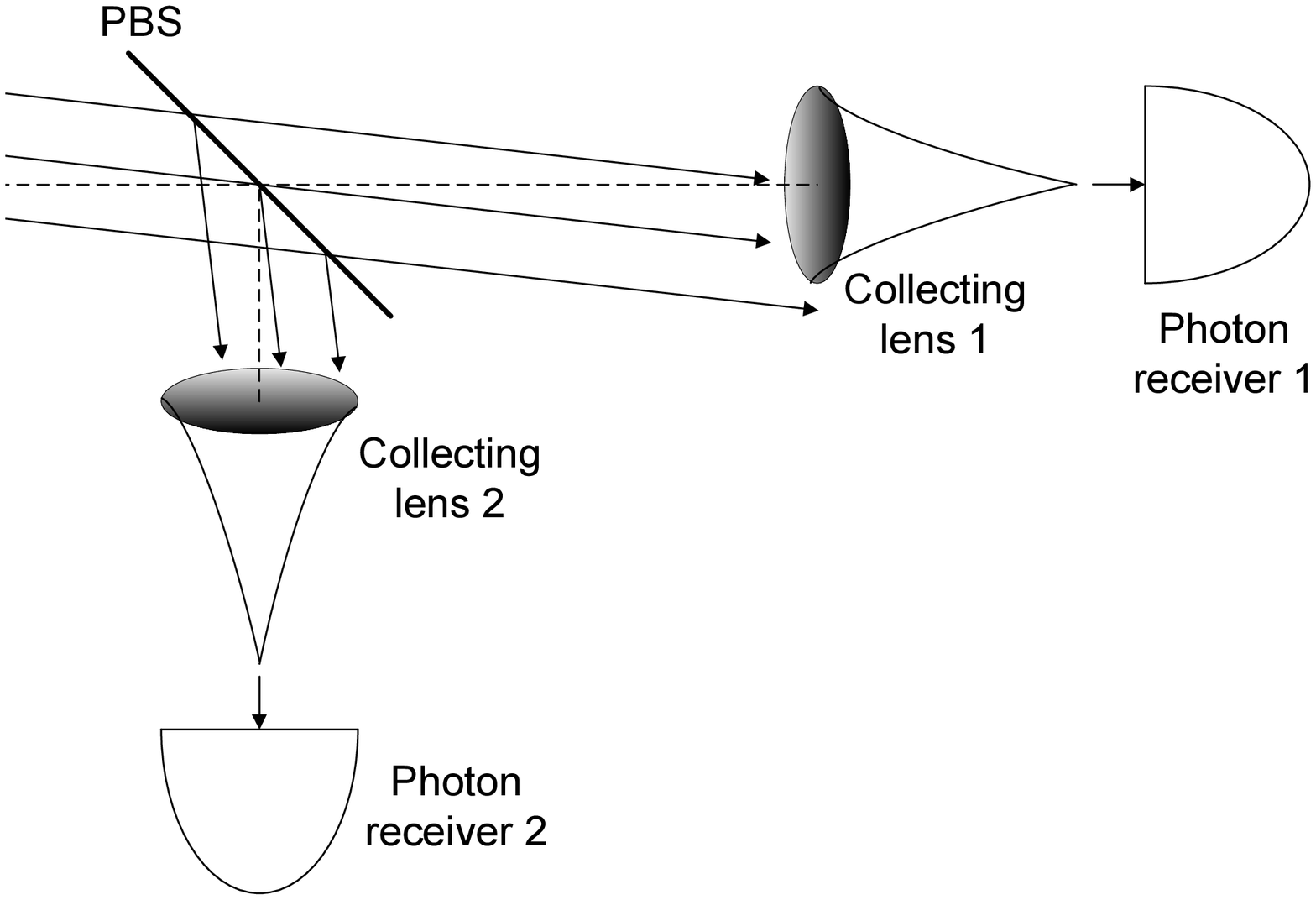}
\end{center}
\fcaption{\label{fig:mismatch-setup2}
Eve changes the incident angle of the input laser beam.
Since the distance between collecting lens 1 and the beamsplitter is not the same as the distance between collecting lens 2 and the beamsplitter, the laser beam arrives at different lateral displacements on the surfaces of the collecting lenses, causing different losses and in turn a mismatch in the detection efficiencies.
Thus, because of this difference in the detection probabilities of bit ``0'' and bit ``1'', Eve may learn some information about the key if Alice and Bob are not aware of this mismatch.
}
\end{figure}

\noindent
The space-domain efficiency mismatch is related to free-space QKD systems \cite{Buttler1998a,Buttler1998b,Buttler2000,Hughes2002,Peng2005,Marcikic2006,Ursin2006,Schmitt-Manderbach2007} where
a change in the spatial mode of the input light may affect the efficiencies of the two detection  channel differently.
In order to illustrate this concept, we show in Fig.~\ref{fig:mismatch-setup2} a simple detection setup.
Here, the two output lights of the beamsplitter pass through the optical coupling systems before being detected.
In practice, there exists asymmetry between the two free-space to fiber coupling systems.
For example, the distances between the coupling systems and the beamsplitter may not be identical and also the lenses of the coupling systems may not be perfectly aligned with respect to the beamsplitter.
If Eve changes the spatial mode of the input laser beam (e.g., angle, lateral displacement), the losses 
in the two coupling systems can potentially be different, leading to detection efficiency mismatch which Eve can take advantage of.
Let's consider an extreme example.
Suppose that the two single-photon detectors (SPD's) are identical, but the distance from the collecting lens of channel 1 to the beamsplitter is twice the distance from the collecting lens of channel 2 to the beamsplitter (see Fig.~\ref{fig:mismatch-setup2}).
In this case, if Eve changes the angle of the input laser beam, the resulting lateral displacement of the laser beam at the surface of the lens of channel 1 is twice that at the surface of the lens of channel 2.
Thus, this induces higher efficiency for channel 2 than for channel 1.
In free-space QKD setups, four detectors are often used with passive basis selection.
In this case, it is conceivable that Eve manipulates the spatial mode of each signal and thereby
produces a bias towards one basis use.

In practice, the process of free-space optical alignment is quite complicated, with
many factors contributing to the coupling efficiency.
Also, errors in the manufacturing process of optical elements can lead to variations in the coupling efficiency.
Spatial change to the input laser beam can further enhance the variations.
Thus, it is expected that a spatial attack can be used to effectively create efficiency mismatch.
Also, note that coupling efficiency mismatch does not vary with time, which liberates Eve from dealing with time-dependent issues.

\subsection{
Frequency-domain detection efficiency mismatch
}
\noindent
In a similar way, detectors may respond to different wavelengths with different efficiencies.
Thus, in principle, Eve may shift the frequency of the incoming signals at Bob to launch her attack.
An acousto-optic modulator can be used to shift the frequency of the input signal up to a few GHz.
On the other hand, by employing nonlinear optical materials, the wavelength of light can be shifted by a few hundreds nm \cite{Albota2004}.

Wavelength filters can be used to ensure that only 
photons within certain spectral band
are permitted to reach the SPD's.
However, one must be careful about the placements of the filters.
If a separate filter is placed before each SPD, Eve may still be able to exploit some efficiency mismatch,
since
the spectral responses of the two wavelength filters may not fully overlap
and thus shifting in frequency may still lead to different efficiencies.
Thus, one simple counter measure is to place one wavelength filter at the entrance of Bob's system.
Conventional SPD's based on Si-APD or InGaAs-APD have a spectral response range of a few hundreds nanometer, while the bandwidth of a narrow-band filter may be 1 nm.
Thus, the responses of the SPD's may be regarded as identical.
On the other hand, other types of SPD's, such as the up-conversion SPD's, may have a bandwidth 
less than 1 nm due to the requirement of phase matching in the up-conversion process.
In this case, a wavelength filter with a bandwidth of 1 nm may not be able to eliminate the efficiency mismatch and
the SPD's may still respond differently to photons with different wavelengths (especially 
when the wavelengths of the input photons get close to the edge of the spectral response window);
thus, a spectral attack may still be possible.

\section{A general detector model\label{sec-model}}
\noindent
\begin{figure}
\includegraphics[width=1\columnwidth]{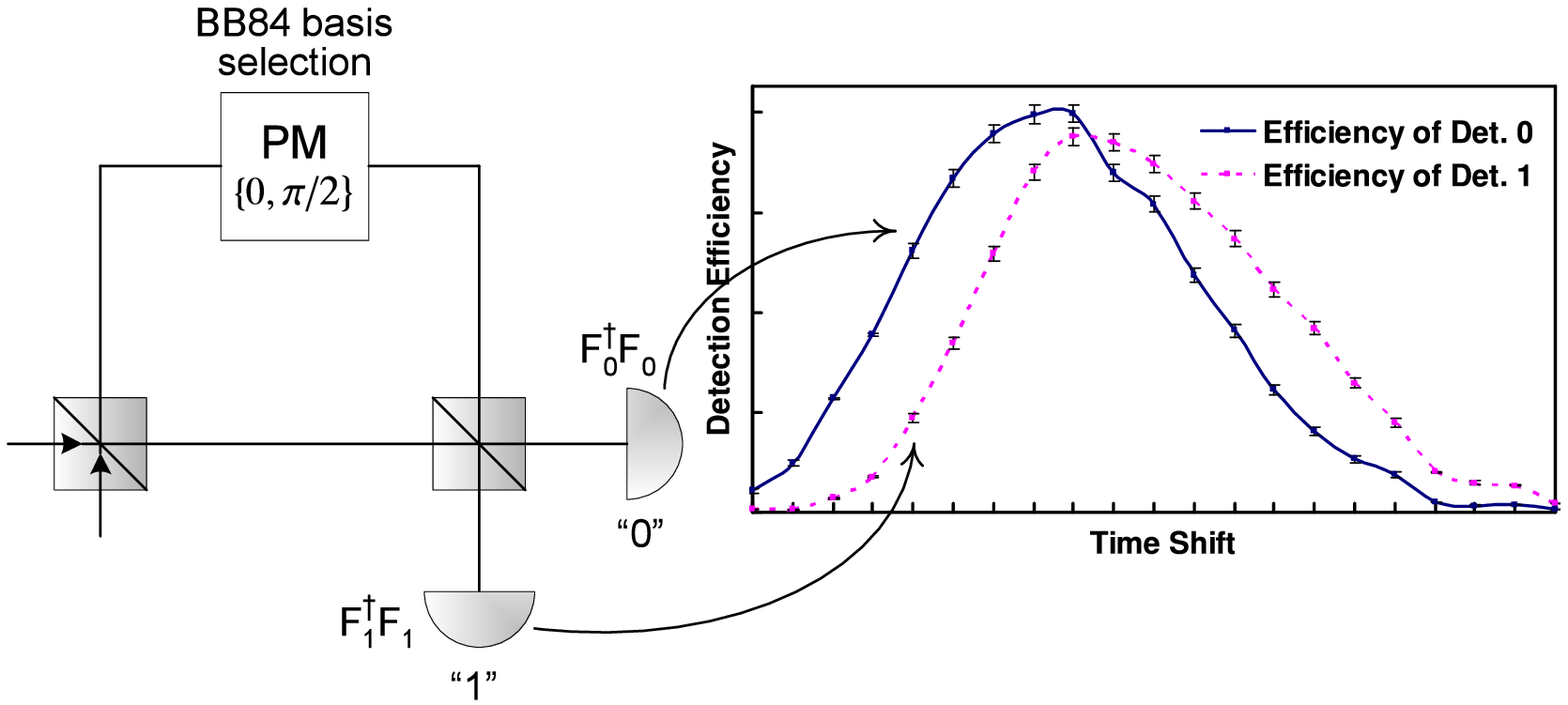}
\fcaption{\label{fig:TwoDetectors}
There are two detectors in the QKD system, one for detecting bit ``0'' and the other for detecting bit ``1''.
The two detectors have different efficiencies and their efficiency responses are characterized by matrices ${F_0}^\dagger F_0$ and ${F_1}^\dagger F_1$.
In this example, the efficiencies depend on the arrival time of the incoming signals, and
the diagonal elements of ${F_i}^\dagger F_i$ expressed in the basis that represents the arrival times are plotted in the figure on the right.
The data of this figure comes from an actual experiment that we performed to demonstrate the time-shift attack \cite{Zhao2007a}.
}
\end{figure}%
In this section, we describe a model for detectors in a QKD system that is general enough to cover a wide variety of efficiency dependencies.
Our security proof 
with efficiency mismatch taken into account will be built on this detector model.
Fig.~\ref{fig:TwoDetectors} shows a QKD system with two detectors having efficiencies dependent on 
some auxiliary dimension (which, in this case, is time).
We are interested in constructing a POVM (positive-operator-valued measurement) for Bob's detector package for detecting bit ``0'' and bit ``1'', taking into account the imperfect efficiencies.
Bob's detector package accepts two systems as input: system $B$ for the qubit representing the information-carrying qubit state of the QKD protocol, and system $T$ representing 
the auxiliary domain (e.g. time, space, and frequency) related to the detection efficiency.

Before we begin the construction of the POVM of Bob's detector package, we state a few 
assumptions.
We assume in the proof that Eve always sends single-photon signals to Bob (see also Sec.~\ref{sec-multi-photon-input} for handling multi-photon input signals).
Furthermore, she sends Bob a qubit state in 
system $B$. 
Dark counts in the detectors may be modeled by Eve sending random qubits.
In addition, we assume that 
in the detector model there is no coupling between the information-carrying qubit system and the auxiliary system.
This means that in the detection process, 
information decoding operations (such as beam splitting) on the qubit system is not affected by the auxiliary system.
This model is not the most general one but is consistent with the three detection scenarios we described in Sec.~\ref{sec-attacks}.
This can be seen by noting that information bits ``0'' and ``1'' (encoded in polarization or phase) are discerned by a beam splitter separating the incoming signal into two paths, and
the beam splitter acts independently
of the arrival time, frequency, or spatial mode of the signal.
Thus, we can assume that the POVM elements take a tensor product form in the qubit system and the auxiliary system.

We now characterize the POVM of the whole detector package consisting of two detectors.
Let's first focus on the $Z$-basis measurement by Bob.
In this case, the POVM elements for measuring bit ``0'' and bit ``1''
can be represented by
\begin{eqnarray}
\label{eqn-general-detector-povm0}
	M_0 &=& \ket{0_z}_B \bra{0_z} \otimes (F_0^\dagger F_0)_T\\
	M_1 &=& \ket{1_z}_B \bra{1_z} \otimes (F_1^\dagger F_1)_T
\end{eqnarray}
where
${F_i}^\dagger F_i$ 
represents the efficiency matrix of detector $i$ that we will discuss in the following.
Thus, the POVM of Bob's measurement is $\{M_0,M_1,\mathbb{I}-M_0-M_1\}$, where the last element represents the case of not getting a click by Bob.
In the special case that the two detectors have constant efficiencies, 
${F_i}^\dagger F_i=\eta_i$ becomes a scalar and Bob's POVM elements for conclusive events become $M_i=\eta_i \ket{i_z}_B \bra{i_z},i=0,1$.
For the later use, it is convenient to 
express Bob's measurement as a filtering operation followed by a simple measurement.
The filtering is
\begin{eqnarray}
	\label{eqn-filtering1}
	F_z &=& \ket{0_z}_B \bra{0_z} \otimes ({F_0})_T + \ket{1_z}_B \bra{1_z} \otimes ({F_1})_T
\end{eqnarray}
and it is followed by the measurement $\{\ket{0_z}_B \bra{0_z} \otimes \mathbb{I}_T,\ket{1_z}_B \bra{1_z} \otimes \mathbb{I}_T\}$.
In a similar way, Bob's $X$-basis measurement can be defined as the filtering operation
\begin{eqnarray}
\label{eqn-filtering-Fx}
F_x&=&\ket{+}_B\bra{+} \otimes ({F_0})_T + \ket{-}_B\bra{-} \otimes ({F_1})_T, 
\end{eqnarray}
followed by the measurement $\{\ket{+}_B \bra{+} \otimes \mathbb{I}_T,\ket{-}_B \bra{-} \otimes \mathbb{I}_T\}$.
In Eqs.~\eqref{eqn-general-detector-povm0}-\eqref{eqn-filtering-Fx}, we
have ignored the possibility of double click events, in view of
their low probabilities.
However, as noted in ILM~\cite{Inamori2005} and GLLP~\cite{Gottesman2004},
one should assign random probabilities to double click events.

The POVM corresponding to detector $i$ consists
of two outcomes $\{{F_i}^\dagger F_i, I-{F_i}^\dagger F_i\}$, where the first (second) element corresponds to having (not having) a detection.
Thus, when the input to detector $i$ is $\rho$ in the auxiliary domain, the probability of detection is $\text{Tr}(\rho {F_i}^\dagger F_i)$.
In general, the efficiency response ${F_i}^\dagger F_i$ is a full matrix:
\begin{eqnarray}
	{F_i}^\dagger F_i
	&=& 
	\begin{bmatrix}
	\eta_i(t_1,t_1) & \cdots & \eta_i(t_1,t_d)\\
	\vdots & & \vdots\\
	\eta_i(t_d,t_1) & \cdots & \eta_i(t_d,t_d)
	\end{bmatrix} .
\end{eqnarray}
This is because there may be a natural or convenient basis to represent it
(e.g. $F_i$ may be expressed in the basis in which one tests the detectors).
Here, we assume that the auxiliary domain has a finite dimension $d$ and it is known to Alice and Bob.
Also, note that the two detectors may not be diagonalizable in the same basis.

One example of the auxiliary domain is the arrival time of the signal (see Fig.~\ref{fig:TwoDetectors}).
The efficiency of gated single-photon detectors can be sensitive to the arrival time of the signals 
and the efficiency response ${F_i}^\dagger F_i$ is conveniently represented in the basis of arrival times.
Thus, one may regard the diagonal term $\eta_i(t_j,t_j)$ as the efficiency of detector $i$ at time $t_j$, and the dimension $d$ is the number of allowable time shifts.

In many cases, one may completely characterize the detectors' responses $F_i$ (with both diagonal and off-diagonal terms).
For example, when the auxiliary domain is polarization, the dimensions of $F_i$ is $2 \times 2$, and its elements can be found by sending signals to the detector with vertical polarization, horizontal polarization, and the in-phase and out-of-phase superpositions of vertical and horizontal polarizations.
On the other hand, when the auxiliary domain is time, there would be infinitely many uncountable time shifts and the dimensions of $F_i$ would be infinite.
Thus, it may appear that perfect characterization is difficult.
However, with a small adjustment, the case of time-dependent efficiency can still be treated with finite dimensions, as we discuss next.

\subsection{Time-dependent efficiency}
\noindent

\subsubsection{Characterization with a finite number of samples}
\noindent
Let us consider that the detectors have an efficiency dependent on the arrival time of the signals.
This is the case for 
detectors operating in gated Geiger mode (e.g., see our experimental time-shift attack paper \cite{Zhao2007a} for how practical efficiency mismatch is exploited by Eve).
When time is the auxiliary dimension, the efficiency response matrix ${F_i}^\dagger F_i$ becomes continuous and contains uncountable elements.
This may be problematic to our analysis.
However, with an additional minor assumption, this problem can be resolved and we can characterize ${F_i}^\dagger F_i$ with a finite number of elements.

The key assumption is that a narrow-band Gaussian-shaped frequency filter is installed at the entrance of Bob, 
with the center frequency matching that of the quantum signal.
Thus, the incoming signals are filtered before reaching the detectors.
The main idea is that 
the signal after the narrow-band filter must also be narrow-band.
Thus, according to the Nyquist-Shannon sampling theorem \cite{Oppenheim1996}, a narrow-band signal can be fully represented by its discrete samples.
This means that the input to the detector 
can always be represented by a series of pulses located at fixed time instants,
and thus 
this allows us to 
characterize the efficiency response of the detector only at those fixed time instants.
Furthermore, when we consider the gating window to have a fixed length, we can characterize the detector with a finite number of pulses sent within the gating window.
Therefore, with an addition of a frequency filter, the characterization of the time-dependent efficiency of the detectors becomes discrete and finite-dimensional, and thus is readily applicable to our security proof.
The details are analyzed in Appendix~\ref{app-time-finite}.
Note that the use of a Gaussian-shaped frequency filter facilitates the use of Gaussian pulses to test the detectors, and Gaussian pulses are easy to generate in practice.
By the same token, adding a time filter makes the characterization of frequency-dependent efficiency discrete.

We remark that the detector and its efficiency response ${F_i}^\dagger F_i$ do not change with time.
What is time dependent is that the detector's efficiency depends on the arrival times of the input signals relative to the detector trigger in a gating window.

\subsubsection{Practical setup for characterization}
\begin{figure}
\includegraphics[width=1\columnwidth]{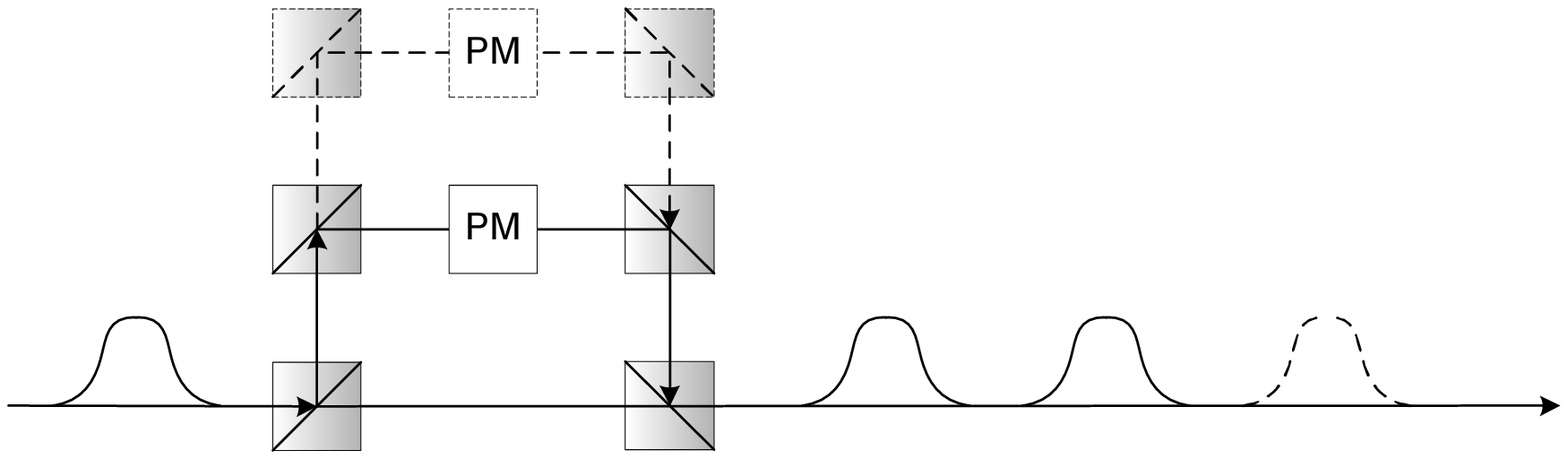}
\fcaption{\label{fig:TestSetup1}
An input signal can be split into more than one pulse by using beamsplitters.
The phase modulators control the the relative phases between these multiple pulses.
One may repeat this setup many times to get a superposition of many pulses.
}
\end{figure}

\noindent
Here, we discuss how to test the time-dependent efficiencies of a detector, denoted by ${F_i}^\dagger F_i$.
Suppose that we choose a basis for ${F_i}^\dagger F_i$ such that its diagonal elements represent the arrival times of the incoming signal.
In this basis, the diagonal elements can be tested easily by sending signals into the detector at different times.
The separation between two adjacent test times is determined by the bandwidth of the frequency filter as discussed above, and the test pulse shape should ideally be Gaussian with a width also determined by the bandwidth of the filter.
The off-diagonal elements can be found by sending in signals in superpositions of the test times with different phases.
One possible practical setup to generate superpositions of the test times is shown in Fig.~\ref{fig:TestSetup1}.
Ideally, one should test the detector using a single-photon source, since this 
is consistent with 
the assumption we use in this paper that Eve always sends single-photon signals to Bob.
In principle, the off-diagonal terms may not be zero, since it is conceivable that 
there exist multiple reflection sites in the detection channel that can give rise to correlations between pulses of different arrival times.
However, in practice, we speculate that the multi-reflected signals are much weaker than the original signals, and therefore the off-diagonal terms may be negligible compared to the diagonal terms.
It would be interesting for future study to test practical detectors for the existence of this correlation.

\section{Security proof for the noiseless case\label{sec-noiseless}}
\noindent
In this section, we prove the security of the case where Eve does not introduce any bit or phase errors but only intervenes with the auxiliary dimension.
Since we assume Alice uses a single-photon source,
the initial state prepared by Alice is 
\begin{eqnarray}
(\ket{00}+\ket{11})_{AB}
.
\end{eqnarray}
Eve does not introduce any noise and she simply attaches an extra system $T$ that represents 
her intervention in the auxiliary dimension, giving
\begin{eqnarray}
\longrightarrow (\ket{00}+\ket{11})_{AB} \otimes \ket{\gamma}_{TE}
.
\end{eqnarray}
Note that the state in $T$ is in general mixed, and thus is purified with system $E$ in this representation.
Now, the state in $BT$ is sent to Bob and he performs the filtering in \myeqnref{eqn-filtering1} to get
\begin{eqnarray}
\longrightarrow  \ket{\Psi_1}&=&F_{z} \left[(\ket{00}+\ket{11})_{AB} \otimes \ket{\gamma}_{TE}\right] 
\nonumber \\
&=& \ket{+}_A \left[ \ket{0}_B F_0 \ket{\gamma}_{TE} + \ket{1}_B F_1 \ket{\gamma}_{TE} \right] 
+ 	\label{eqn-state-after-Fz} \\
&& \ket{-}_A \left[ \ket{0}_B F_0 \ket{\gamma}_{TE} - \ket{1}_B F_1 \ket{\gamma}_{TE} \right] 
 \nonumber
\end{eqnarray}
where $F_i$ acts on system $T$.
In the actual protocol, after the filtering operation $F_{z}$,
both Alice and Bob perform measurements on their systems in the $Z$ basis to obtain their raw keys.
Bob corrects the bit errors in his raw key according to the error syndromes sent by Alice through a secret classical channel, thus making Alice's and Bob's raw keys the same.
After this error correction step, they multiply the same random matrix to their raw keys (for privacy amplification) to arrive at the final secret key.
The essence of Koashi's proof \cite{Koashi2005b} is that since the final secret key is derived from Alice's raw key obtained in the $Z$ basis, Alice and Bob's main goal is to guarantee that system $A$ of Alice is in an $X$ eigenstate.
In this case, because the uncertainty of system $A$ in the $X$ basis is minimized, by the uncertainty principle, Eve's uncertainty of system $A$ in the $Z$ basis is maximized, and thus the final key is secret to Eve.

\begin{figure}
\subfigure[]{
\label{fig:measurement1-top}
\centerline{
\includegraphics[width=.9\columnwidth]{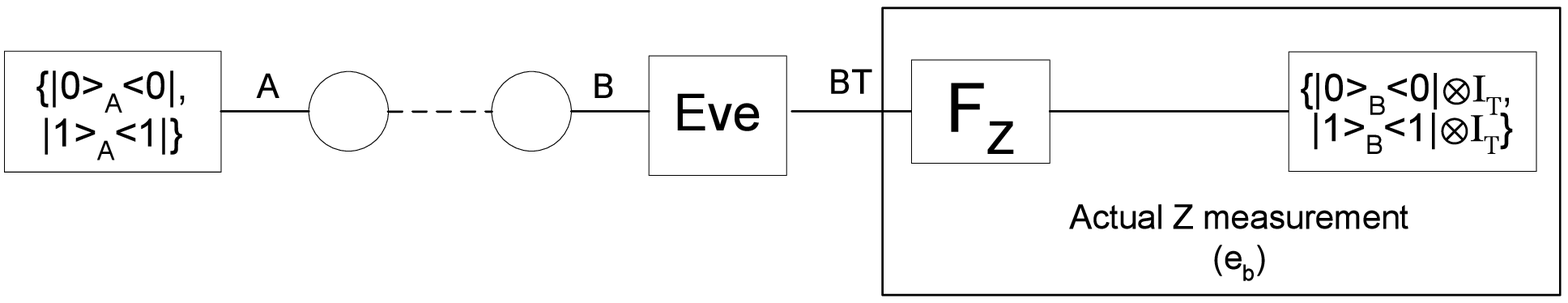}}
}
\\
\subfigure[]{
\label{fig:measurement1-middle}
\centerline{
\includegraphics[width=.9\columnwidth]{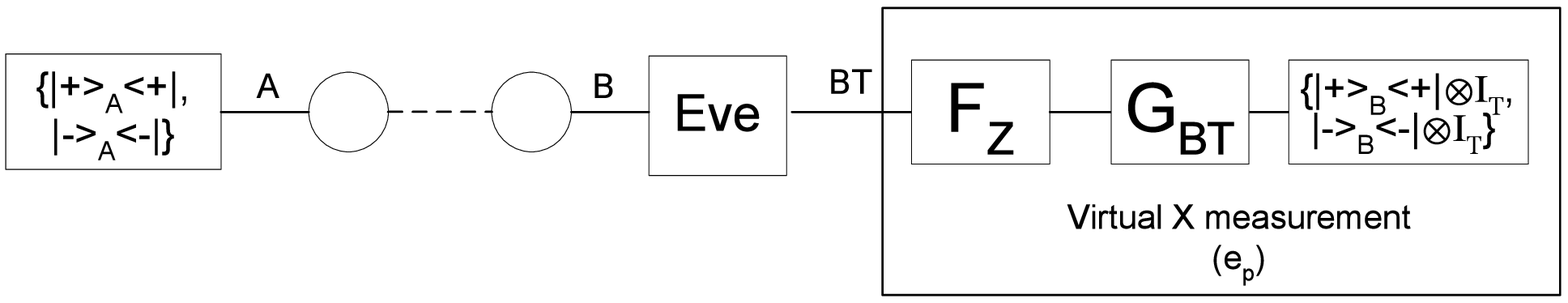}}
}
\\
\subfigure[]{
\label{fig:measurement1-bottom}
\centerline{
\includegraphics[width=.9\columnwidth]{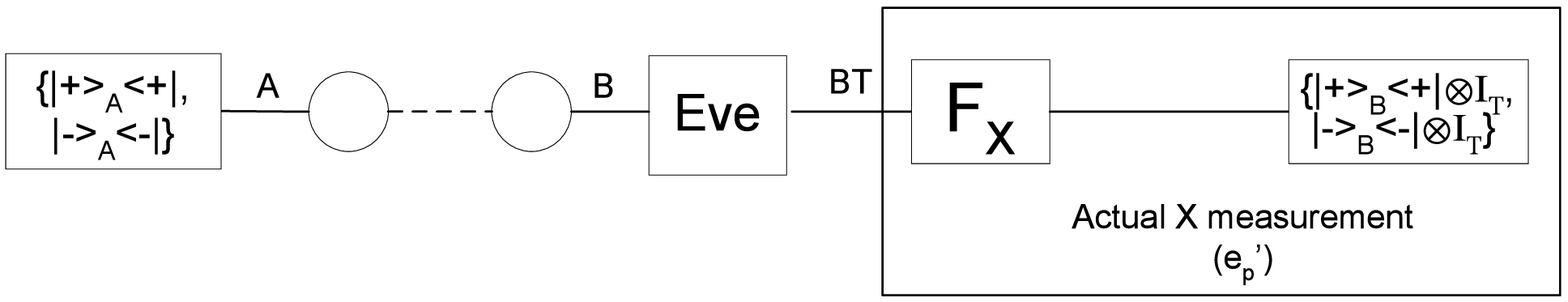}}
}
\fcaption{\label{fig:measurement1}
The top figure shows the actual measurement by Bob to general key bits in the $Z$-basis, with the corresponding bit error rate of $e_b$.
The middle figure shows the virtual measurement by Bob for helping Alice predict her $X$-basis measurement outcomes on system $A$ for the key bits generated from $Z$-basis measurements, with the corresponding (virtual) phase error rate of $e_p$.
The bottom figure shows the actual measurement by Bob in the $X$-basis, with the corresponding (actual) phase error rate of $e_p'$.
}
\end{figure}

Thus, to prove security using Koashi's proof,
we consider that
Bob performs a virtual measurement on system $BT$ in order to predict Alice's $X$-basis measurement outcome on system $A$ (see Fig.~\ref{fig:measurement1-middle}).
Essentially, system $T$ (though not carrying bit information) acts as a shield \cite{Horodecki2005} that protects Alice and Bob's key from Eve.
Loosely speaking, the uncertainty associated with using Bob's virtual measurement to predict Alice's $X$-basis measurement outcome is related to Eve's information on the key before privacy amplification.
The advantage of using Koashi's proof is that there is no restriction on Bob's virtual measurement on system $BT$.
He is allowed to perform any quantum measurement after the filter $F_{z}$.
This measurement is not performed in practice and thus whether it is physically realizable is not a concern.
The only concern (or restriction) is that we have to make sure that statistics
of virtual protocol can be well estimated using those of the actual protocol.
Thus, we would like to construct a virtual measurement for Bob aiming at predicting Alice's $X$-basis measurement outcome with high certainty.

\subsection*{Procrustean method of filtering for Bob's virtual measurement}
\noindent
The objective for Bob is to distinguish the two non-orthogonal states in system $BT$ corresponding to Alice's $\ket{+}$ and $\ket{-}$ states in \myeqnref{eqn-state-after-Fz}.
The non-orthogonality stems from the efficiency mismatch between the two detectors and this makes the problem more difficult than the perfect-detector case.
What complicates the problem even further is that the state Bob tries to measure is not known a priori, owing to the fact that the initial state in system $T$ is unknown.
Nevertheless, we propose a simple way to identify the two non-orthogonal states, which is by orthogonalizing them with a filter that may succeed with a probability less than one.
The idea of orthogonalizing each signal independently is known as the Procrustean method of filtering \cite{Bennett1996}.
Although our filtering method may not be optimal since we are operating on individual signals independently, it does provide a simple and intuitive method 
that magically orthogonalizes the two states even though the initial state in system $T$ is unknown.

In this paper, we specialize
in a trick
to constructing such a virtual measurement 
by inverting the filtering operations $F_i$ in Eq.~\eqref{eqn-state-after-Fz} as follows.
Bob first performs a filtering 
\begin{eqnarray}
\label{eqn-BobVirtualFilter}
	G_{BT} &=&
	\ket{0_z}_B \bra{0_z} \otimes (C F_0^{-1})_T + \ket{1_z}_B \bra{1_z} \otimes  (C F_1^{-1})_T	
\end{eqnarray}
on system $BT$ of Eq.~\eqref{eqn-state-after-Fz} and then performs an $X$-basis measurement on system $B$
(i.e., with POVM 
$\{\ket{+}_B \bra{+} \otimes \mathbb{I}_T,\ket{-}_B \bra{-} \otimes \mathbb{I}_T\}$).
Here, we introduce a $d\times d$ matrix $C$ to ensure that $G_{BT}$ is a valid filtering operation (i.e., $C$ is chosen so that $G_{BT}^\dagger G_{BT} \leq \mathbb{I}$).
See Appendix~\ref{app-virtual-filter} for the derivation of $C$.

After applying this filter $G_{BT}$ to Eq.~\eqref{eqn-state-after-Fz}, the whole state becomes
\begin{eqnarray}
\longrightarrow && G_{BT} \ket{\Psi_1} \nonumber \\
&=&[\ket{++}_{AB}+\ket{--}_{AB}] \otimes C \ket{\gamma}_{TE} 
. \label{eqn-state-after-GBT}
\end{eqnarray}
where $C$ acts on system $T$.
The magic of the filter $G_{BT}$ is that even though the initial state of system $T$ is unknown, the filter is still able to concentrate an perfect EPR pair in system $AB$.
Since now there is an EPR pair in system $AB$,
an $X$-basis measurement outcome on system $B$ is perfectly correlated with an $X$-basis measurement outcome on system $A$.
Thus, when Bob measures in the $X$ basis on system $B$ of the successfully filtered state in Eq.~\eqref{eqn-state-after-GBT}, he is able to predict Alice's $X$-basis measurement outcome with complete certainty, and this is enough to prove security.
Once Alice and Bob know that Alice's state is an $X$ eigenstate, the final key derived from the $Z$-basis measurements of Alice's state is then secret to Eve.
The probability of successful filtering is 
\begin{eqnarray}
	p_\text{succ}
	&=&
	\frac{Tr[G_{BT} \ket{\Psi_1} \bra{\Psi_1} G_{BT}^\dagger]}{Tr[\ket{\Psi_1} \bra{\Psi_1}]} \\
	&=&
	\frac{2\bra{\gamma} (C^\dagger C)_T \otimes I_E  \ket{\gamma}_{TE}}{\bra{\gamma} (F_0^\dagger F_0 + F_1^\dagger F_1)_T \otimes I_E \ket{\gamma}_{TE}} . \label{eqn-psucc1}
\end{eqnarray}
Therefore, among all the $N$ qubit pairs shared by Alice and Bob,
Bob is able to predict Alice's $X$-basis measurement outcomes of 
$N p_\text{succ}$
key bits, 
since he knows whether the virtual filtering on each bit succeeded or not when it is performed. 
By applying the arguments of Koashi's proof, 
the secret key generation rate is simply $p_\text{succ}$.
Since the state $\ket{\gamma}$ 
in \myeqnref{eqn-psucc1} is chosen by Eve,
the worst-case final secret key generation rate on detected signals is
\begin{equation}
\label{eqn-noiseless1}
R_\text{noiseless} = \max_C \min_{\ket{\gamma}} p_\text{succ},
\end{equation}
which is a lower bound on the key generation rate.
The detectors' efficiency matrices $F_0^\dagger F_0$ and $F_1^\dagger F_1$ are assumed to be known.
In this case, the final secret key generation rate on detected signals is found by solving \myeqnref{eqn-noiseless1} (see Appendix~\ref{app-virtual-filter} for detail) and it is equal to
\begin{equation}
\label{eqn-noiseless2}
R_\text{noiseless}
=
\frac{2}{1+\max\left(D_1,\frac{1}{D_1},\ldots,D_d,\frac{1}{D_d}\right)}
\end{equation}
where
$D=\text{diag}(D_1,\ldots,D_d)$ is a diagonal matrix with positive real elements (which are the eigenvalues) and is  determined from the Hermitian decomposition of $F_0 (F_1^\dagger F_1)^{-1} F_0^\dagger = U D U^\dagger$.
The elements of $D$ represents the ratios of the efficiencies of the two detectors and when the assignment of $F_0$ and $F_1$ is reversed, $D$ would also be inverted.
If the efficiency matrices $F_i$ are only partially known, one can find the worst-case key generation rate by also minimizing \myeqnref{eqn-noiseless1} over them.
Throughout the derivation, we have assumed that $F_0$ and $F_1$ have full rank (or invertible).
Otherwise, we can show that the key generation rate is zero (see below).
Thus, characterization of the detectors is very important.

\subsection*{Special case 1: $F_i$ is not invertible}
\noindent
Suppose that $F_0$ is not invertible (i.e., not full rank) and the nullspace of $F_0$ and that of $F_1$ are different\footnote{If $F_0$ and $F_1$ have the same nullspaces, then they can be reduced to two invertible matrices.} ($F_1$ may be invertible).
In this case, the worst-case final key generation rate is zero.
One may observe this from \myeqnref{eqn-state-after-Fz}.
Eve may choose $\ket{\gamma}$ to be in the nullspace of $F_0$ and not in the nullspace of $F_1$, leading to
system $A$ completely disentangled with system $B$.
Thus, no key can be generated in this case, since entanglement is a precondition for generating secret keys \cite{Curty2004}.

\subsection*{Special case 2: only diagonal terms are known}
\noindent
Note that if only the diagonal terms of $F_0^\dagger F_0$ and $F_1^\dagger F_1$ are known, the worst-case final key generation rate is zero.
Since only the diagonal terms are known, the final key generation rate is determined by minimizing over the off-diagonal terms.
One can imagine that the off-diagonal terms are chosen such that $F_0$ is not invertible and $F_1$ is invertible.
Thus, this case is reduced to special case 1, which allows us to conclude that the final key generation rate in this case is zero.

\subsection*{Special case 3: $F_0$ and $F_1$ are diagonal}
\noindent
Here, we consider the special case that the Bob's detectors' responses to the time-shifts, ${F_i}^\dagger F_i, i=0,1$, are diagonal, i.e., there is no correlation in the efficiencies between time-shifts.
Suppose the efficiency matrices are ${F_i}^\dagger F_i = \text{diag}(\eta_i(t_1),\eta_i(t_2),\ldots)$.
We compute the final key generation rate using \myeqnref{eqn-noiseless2}.
Note that $F_0 (F_1^\dagger F_1)^{-1} F_0^\dagger = \text{diag}\left(\frac{\eta_0(t_1)}{\eta_1(t_1)},\frac{\eta_0(t_2)}{\eta_1(t_2)},\ldots\right)$.
Using this fact, the final key generation rate on detected signals is
\begin{equation}
\label{eqn-noiseless3}
R_\text{noiseless, diag}
=
\min_t \frac{2 \min( \eta_0(t), \eta_1(t) ) }{\eta_0(t) + \eta_1(t)} .
\end{equation}

\section{General security proof
\label{sec-noisy}}
\noindent
In this section, we prove security under the most general attack by Eve in which she can coherently process the signals sent by Alice and perform a joint measurement on her ancillas.
In this general case,
Eve's action on the $l$th bit can be described by her preparing a pure state system $T$ (the auxiliary domain to which the detectors respond) and performing a superoperation on both systems $B$ and $T$, as follows:
\begin{eqnarray}
\label{eqn-noisy-state-after-Eve}
\ket{\Psi_2^{(l)}} &=&
\sum_i E_{BT}^{(l,i)} 
(\ket{00}+\ket{11})_{AB} \otimes \ket{0}_{T}
\otimes \ket{i}_E
\end{eqnarray}
where $E_{BT}^{(l,i)}$ is the operation element for the $l$th bit and it is responsible for introducing bit and phase errors.
Note that it can depend on Eve's action on all other bits.
Now, the state in $BT$ is sent to Bob and he performs the filtering in \myeqnref{eqn-filtering1} as in the noiseless case to get
\begin{eqnarray}
\label{eqn-noisy-pre-virtual-filtered-state1}
\longrightarrow  
F_{z} \ket{\Psi_2^{(l)}} .
\end{eqnarray}
As in 
Fig.~\ref{fig:measurement1-middle}, in order to estimate the amount of privacy amplification, we assume that Bob performs the virtual measurement by 
applying 
a filter to $F_z \ket{\Psi_2^{(l)}}$.
Bob applies the same filter 
$G_{BT}$ in \myeqnref{eqn-BobVirtualFilter} as in the noiseless case
 and we have
\begin{eqnarray}
\longrightarrow  \ket{\Psi_3^{(l)}} &=& G_{BT} F_z \ket{\Psi_2^{(l)}} .
\label{eqn-state-after-GBT-noisy}
\end{eqnarray}
Thus, after Bob measures in the $X$ basis on system $B$ of the successfully filtered state in Eq.~\eqref{eqn-state-after-GBT-noisy}, he is able to predict Alice's $X$-basis measurement outcome with uncertainty indicated by the virtual phase error probability of $\ket{\Psi_3^{(l)}}$.
On the other hand, the actual phase error probability ($e_p'$) is not generated by this state; it is generated by some other state, namely $F_x \ket{\Psi_2^{(l)}}$ [
 Fig.~\ref{fig:measurement1-bottom}].
Thus, we need to estimate the 
virtual phase error probability ($e_p$) of the filtered state $\ket{\Psi_3^{(l)}}$ [
Fig.~\ref{fig:measurement1-middle}] 
from the actual phase error probability ($e_p'$)
[
Fig.~\ref{fig:measurement1-bottom}].
In addition, we need to lower bound the probability of successful filtering of the state $F_z \ket{\Psi_2^{(l)}}$.
Overall, we are interested in the formulas for
the actual bit error probability ($e_b$) of $F_z \ket{\Psi_2^{(l)}}$, the virtual phase error probability ($e_p$) of $\ket{\Psi_3^{(l)}}$, the actual phase error probability ($e_p'$) of $F_x \ket{\Psi_2^{(l)}}$, and the virtual filtering probability ($p_\text{succ,noisy}$) of $F_z \ket{\Psi_2^{(l)}}$:
\begin{eqnarray}
e_b
&=&
\frac
{
\sum_{l}
\text{Tr}[ P_\text{bit} F_z \ket{\Psi_2^{(l)}} \bra{\Psi_2^{(l)}} F_z^\dagger]}
{
\sum_{l}
\bra{\Psi_2^{(l)}} F_z^\dagger F_z \ket{\Psi_2^{(l)}} }
\label{eqn-noisy-eb1}
\\
e_p
&=&
\frac
{\sum_{l}
\text{Tr}[ P_\text{phase} \ket{\Psi_3^{(l)}} \bra{\Psi_3^{(l)}} ]}
{\sum_{l}
\langle\Psi_3^{(l)} \ket{\Psi_3^{(l)}} }
\label{eqn-noisy-ep1}
\\
e_p'
&=&
\frac
{\sum_{l}
\text{Tr}[ P_\text{phase} F_x \ket{\Psi_2^{(l)}} \bra{\Psi_2^{(l)}} F_x^\dagger]}
{\sum_{l}
\bra{\Psi_2^{(l)}} F_x^\dagger F_x \ket{\Psi_2^{(l)}} }
\label{eqn-noisy-epp1}
\\
p_\text{succ,noisy}
&=&
\frac
{\sum_{l}
\langle\Psi_3^{(l)} \ket{\Psi_3^{(l)}} }
{\sum_{l}
\bra{\Psi_2^{(l)}} F_z^\dagger F_z \ket{\Psi_2^{(l)}} }
\label{eqn-noisy-psucc1}
\end{eqnarray}
where $P_\text{bit}=\ket{01}_{AB} \bra{01} +\ket{10}_{AB} \bra{10}$ and $P_\text{phase}=\ket{+-}_{AB} \bra{+-} +\ket{-+}_{AB} \bra{-+}$.
These quantities are specified by Eve through her selection of the attack strategy
 $E_{BT}^{(l,i)}$.
Although there seems to be many dimensions in the attack strategy (as $l$ runs over all qubit pairs and $i$ over any range), one can simplify the attack into a small number of dimensions when probabilities are concerned.

For the case we consider in this paper, it is shown in Appendix~\ref{app-simplification} that the various probabilities of interest in Eqs.~\eqref{eqn-noisy-eb1}-\eqref{eqn-noisy-psucc1} can be simplified with a collective attack to
\begin{eqnarray}
e_b
&=&
\frac
{
\operatorname{Tr}[ \rho_E (\tilde{Z}_{10} \otimes F_0^\dagger F_0 +  \tilde{Z}_{01} \otimes F_1^\dagger F_1 )]
}
{
\operatorname{Tr}[ \rho_E ((\tilde{Z}_{00}+\tilde{Z}_{10}) \otimes F_0^\dagger F_0 +  (\tilde{Z}_{11}+\tilde{Z}_{01}) \otimes F_1^\dagger F_1 )]
}
\label{eqn-noisy-eb2}
\\
e_p
&=&
\frac
{
\operatorname{Tr}[ \rho_E (\tilde{X}_{-+}+\tilde{X}_{+-}) \otimes C^\dagger C]
}
{
\operatorname{Tr}[ \rho_E (\tilde{X}_{++}+\tilde{X}_{-+}+\tilde{X}_{+-}+\tilde{X}_{--}) \otimes C^\dagger C]
}
\label{eqn-noisy-ep2}
\\
e_p'
&=&
\frac
{
\operatorname{Tr}[ \rho_E (\tilde{X}_{-+} \otimes F_0^\dagger F_0 +  \tilde{X}_{+-} \otimes F_1^\dagger F_1 )]
}
{
\operatorname{Tr}[ \rho_E ((\tilde{X}_{++}+\tilde{X}_{-+}) \otimes F_0^\dagger F_0 +  (\tilde{X}_{--}+\tilde{X}_{+-}) \otimes F_1^\dagger F_1 )] 
}
\label{eqn-noisy-epp2}
\\
p_\text{succ,noisy}
&=&
\frac
{
\operatorname{Tr}[ \rho_E ((\tilde{Z}_{00}+\tilde{Z}_{10}+\tilde{Z}_{11}+\tilde{Z}_{01}) \otimes C^\dagger C)]
}
{
\operatorname{Tr}[ \rho_E ((\tilde{Z}_{00}+\tilde{Z}_{10}) \otimes F_0^\dagger F_0 +  (\tilde{Z}_{11}+\tilde{Z}_{01}) \otimes F_1^\dagger F_1 )]
}
\label{eqn-noisy-psucc2}
\end{eqnarray}
where  
$\rho_E$
represents Eve's action, which is
averaged over all signals,
and has dimensions $4d \times 4d$ (again, $d \times d$ are the dimensions of $F_i$ and $C$)
and $C$ is from \myeqnref{eqn-matrixC}.
Here, $\tilde{Z}_{i,j}$ and $\tilde{X}_{i,j}$ are
constant $4 \times 4$ matrices
given in \myeqnref{eqn-4x4-matrices}.
[Note that we are using $C$ determined in the noiseless case.
However, it can be shown (in a similar fashion as in the noiseless case) that the same $C$ is obtained if we maximize the filtering probability $p_\text{succ,noisy}$ over $C$ (which is related to problem \eqref{eqn-noisy-optimize-psucc-uncon} in Section~\ref{sec-subopt-bounds}).]

It is worth noting that the security proofs for a three-state protocol \cite{Boileau2005} and the SARG04 protocol \cite{Tamaki2006,Fung2006} that share the same technique (the Azuma's inequality \cite{Azuma1967}) as the current paper reduce to collective attacks in a more straightforward manner.
In their cases, the normalizations of all probabilities of interest (the bit and phase error probabilities) are the same and this together with the concavity of the relations between the probabilities immediately reduce joint attacks to collective attacks.
In the case of the current paper, the normalizations (i.e., the denominators of  Eqs.~\eqref{eqn-noisy-eb1}-\eqref{eqn-noisy-psucc1}) are different; and thus we need a more involved analysis to reduce to collective attacks.

We remark that in our general method, we use the Procrustean Method
and the trick of inverting the detection efficiency filters when constructing the virtual filter.
It is well-known (e.g. Ref.~\cite{Bennett1996})
that in entanglement distillation theory 
the Procrustean Method is, in general, sub-optimal.
On the other hand, our method, despite its sub-optimality, provides an instructive and easily understandable way to prove security.

\subsection{Bounding filtering probability $p_\text{succ,noisy}$ and phase error probability $e_p$}
\noindent
Both the virtual filtering probability ($p_\text{succ,noisy}$) and virtual phase error probability ($e_p$) ultimately determine the key generation rate (cf. \myeqnref{eqn-noisy-final-key-rate}).
Thus, we bound them by
numerically optimizing 
them
over Eve's action subject to the observed bit and phase error rates.
One important tool that we rely on to identify the the observed rates with the corresponding probabilities is the Azuma's inequality~\cite{Azuma1967}, 
which asserts that the sum of the probabilities for an event over all trials is asymptotically close to the observed count of the event (this inequality was similarly used in other security proofs \cite{Boileau2005,Tamaki2006,Fung2006,Fung2006b,Tamaki2006b})\footnote{
In order to see how the Azuma's inequality is applied to Eqs.~\eqref{eqn-noisy-eb1}-\eqref{eqn-noisy-psucc1},
consider $e_b$ in Eq.~\eqref{eqn-noisy-eb1} as a concrete example.
We apply the Azuma's inequality to the numerator and denominator separately.
The numerator in the right hand side of Eq.~\eqref{eqn-noisy-eb1} is Pr\{bit error and conclusive result\} and the denominator is Pr\{conclusive result\}.
Since $e_b$ is defined as the bit error rate {\em conditional on} conclusive bits, we can apply the Azuma's inequality to the numerator of Eq.~\eqref{eqn-noisy-eb1} to get the actual
number of bits with error and
apply the Azuma's inequality to the denominator to get the
actual number of conclusive bits.
Dividing these two numbers gives us $e_b$.
}.
Thus, the rate of successful virtual filtering 
can be lower bounded as 
\begin{align*}
\text{minimize    } & p_\text{succ,noisy} \nonumber
\\
\text{subject to    } & e_b = \text{observed }e_b \tag{P1} \label{eqn-noisy-optimize-psucc}\\
& e_p' = \text{observed }e_p' \nonumber
\end{align*}
and the virtual phase error rate 
can be upper bounded as 
\begin{align*}
\text{maximize    } & e_p \nonumber
\\
\text{subject to    } & e_b = \text{observed }e_b \tag{P2} \label{eqn-noisy-optimize-ep}\\
& e_p' = \text{observed }e_p' \nonumber
\end{align*}
where the optimization is over Eve's action 
$\rho_E$
and the formulas for the error probabilities and the filtering probability are from Eqs.~\eqref{eqn-noisy-eb2}-\eqref{eqn-noisy-psucc2}.
Note that both problems \eqref{eqn-noisy-optimize-psucc} and \eqref{eqn-noisy-optimize-ep} can be expressed as a polynomial problem by multiplying the denominator of each fraction in each of the constraints; and thus the two problems can be solved numerically and efficiently%
\footnote{\label{footnote-rank1}It seems that 
considering rank-one $\rho_E$
is always sufficient to achieve the optimal solution based on our simulation results.
However, we have not been able to prove this.
Note that convexity/concavity of \eqref{eqn-noisy-optimize-psucc}/\eqref{eqn-noisy-optimize-ep} does not immediately give rise to this conclusion.}
\cite{Lasserre2001,Vandenberghe1996}.

When $e_b=e_p'=0$, solving these two problems gives the same noiseless-case result in \myeqnref{eqn-noiseless2}.

When $F_0=F_1$ (meaning that there is no efficiency mismatch), it can easily be checked that $p_\text{succ,noisy}=1$ and $e_p=e_p'$, as expected.

We remark that although both problems \eqref{eqn-noisy-optimize-psucc} and \eqref{eqn-noisy-optimize-ep} involve only the average error rates,  improvement in the bounds may be obtained by, for example,
separating the error rates for bits ``0'' and ``1''.

\subsection{Suboptimal bounds\label{sec-subopt-bounds}}
\noindent
In order to better understand the relationship between the efficiency mismatch and the phase error probability $e_p$ or the filtering probability $p_\text{succ,noisy}$, we compute suboptimal bounds for these two quantities.
These bounds are obtained by simply dropping the constraints in problems \eqref{eqn-noisy-optimize-psucc} and \eqref{eqn-noisy-optimize-ep}.
Specifically, we solve
\begin{align*}
\text{minimize    } & p_\text{succ,noisy} \tag{P1'} \label{eqn-noisy-optimize-psucc-uncon}
\end{align*}
and
\begin{align*}
\text{maximize    } & \frac{e_p}{e_p'} \tag{P2'} \label{eqn-noisy-optimize-ep-uncon}
\end{align*}
where the optimization is over Eve's action 
$\rho_E$
and the formulas for the error probabilities and the filtering probability are from Eqs.~\eqref{eqn-noisy-eb2}-\eqref{eqn-noisy-psucc2}.
Solving these problems gives (see Appendix~\ref{app-solve-suboptimal} for detail)
\begin{align}
p_\text{succ,noisy} & \geq
\min\left(D_1,\frac{1}{D_1},\ldots,D_d,\frac{1}{D_d}\right) 
\label{eqn-psucc-suboptimal1}
\\
\frac{e_p}{e_p'} & \leq 
\max\left(D_1,\frac{1}{D_1},\ldots,D_d,\frac{1}{D_d}\right)
\label{eqn-ep-suboptimal1}
\end{align}
where
$D=\text{diag}(D_1,\ldots,D_d)$ is a diagonal matrix with positive real elements and is  determined from the Hermitian decomposition of $F_0 (F_1^\dagger F_1)^{-1} F_0^\dagger = U D U^\dagger$.
In other words, the virtual filtering probability and the virtual phase error probability are related to the minimum and maximum efficiency ratios between the two detectors.
Note that the right hand side of \myeqnref{eqn-psucc-suboptimal1} is the inverse of that of \myeqnref{eqn-ep-suboptimal1}.

\subsection{Key generation rate}
\noindent
After the virtual filtering operation, Bob can predict Alice's $X$-basis measurement outcomes with an error probability of $e_p$ for $N p_\text{succ,noisy}$ key bits.
In light of Koashi's proof, 
Bob's prediction removes some uncertainty on Alice's $X$-basis measurement outcomes (through Bob's  communication to Alice).
The remaining uncertainty can be removed (with high probability) by Alice performing hashing on the key bits
in $m$ rounds, where $m$ is less than $N$ in general.
In our case, $m=(N-N p_\text{succ,noisy}) + N p_\text{succ,noisy} H_2(e_p)$ where the first (second) part represents the key bits for which Bob's virtual filter did not (did) succeed.
Here, $H_2(x)=-x\log_2x-(1-x)\log_2(1-x)$ is the binary entropy function.
According to the proof, the number of secure key bits generated is (without error correction)
\begin{eqnarray}
K_\text{PA} &=& N-m \nonumber\\
&=&
N p_\text{succ,noisy} (1-H_2(e_p)) .
\label{eqn-noisy-key-rate1}
\end{eqnarray}
The amount of (pre-shared) secret key bits sacrificed for error correction using encrypted one-way communication is
\begin{align}
K_\text{EC} & = N H_2(e_b) .
\end{align}
Combining them gives the final key generation rate on detected signals:
\begin{eqnarray}
\label{eqn-noisy-key-rate2}
R_\text{noisy} &=& [K_\text{PA} - K_\text{EC}]/N \nonumber\\
&=&
p_\text{succ,noisy} (1-H_2(e_p)) - H_2(e_b)
\label{eqn-noisy-final-key-rate}
\end{eqnarray}
where  $p_\text{succ,noisy}$ and $e_p$ are obtained by solving problems \eqref{eqn-noisy-optimize-psucc} and \eqref{eqn-noisy-optimize-ep}, respectively.
Alternatively, they may be obtained from the suboptimal bounds in Eqs.~\eqref{eqn-psucc-suboptimal1} and \eqref{eqn-ep-suboptimal1}.
Also, it is worth noting that we do not need to separately bound $p_\text{succ,noisy}$ and $e_p$ with two separate optimization problems.
We can instead minimize the key generation rate in \myeqnref{eqn-noisy-final-key-rate} subject to the observed error rates (which are the common constraints of \eqref{eqn-noisy-optimize-psucc} and \eqref{eqn-noisy-optimize-ep}).
However, this problem is not a convex optimization problem (due to the nonlinear equality constraints) or a polynomial optimization problem (due to the non-polynomial objective function) and thus may not easily be solved optimally and efficiently.

\subsection{Example 1: scalar efficiencies\label{subsec-example1}}
\noindent 
Consider a QKD system in which the two detectors have constant but different efficiencies.
This means that the detection efficiency matrices, ${F_i}^\dagger F_i , i=0,1$, are scalar.
In this case, one can easily compute the noiseless key generation rate on detected signals by using \myeqnref{eqn-noiseless2} or \myeqnref{eqn-noiseless3} to be
\begin{equation}
\label{eqn-example-scalar-noiseless}
R_\text{noiseless, scalar}
=
\frac{2 \min( \eta_0, \eta_1 ) }{\eta_0 + \eta_1} ,
\end{equation}%
where $\eta_i={F_i}^\dagger F_i$.
On the other hand, this formula for the scalar efficiency case can be obtained by the following simple argument.
Bob can artificially make the efficiencies of the two detectors the same by
randomly discarding some detection events of the detector with the higher efficiency in the data postprocessing step.
To arrive at the above formula, suppose that Alice sends $N$ signals (with equal numbers of bits ``0'' and ``1'') to Bob.  
Then, Bob will detect $\frac{N}{2} \eta_0$ number of ``0''s and $\frac{N}{2} \eta_1$ number of ``1''s, giving the total number of detected signals $\frac{N}{2} (\eta_0+\eta_1)$.
Now, assuming that $\eta_0<\eta_1$, Bob discards some of detection events of the detector corresponding to bit ``1'', so that 
effectively $\frac{N}{2} \eta_0$ number of ``0''s and $\frac{N}{2} \eta_0$ number of ``1''s are retained to form the final key (there is no error correction or privacy amplification since there is no noise).
Thus, the final key length is $N \eta_0$, and dividing this by the total number of detected signals gives \myeqnref{eqn-example-scalar-noiseless}.

Now, let's consider the noisy case.
The above simple argument with data discarding can also handle the noisy case.
After the artificial equalization of the two detection efficiencies, the situation Alice and Bob are facing becomes that considered in Shor-Preskill \cite{Shor2000}.
This can be seen as follows.
When the efficiencies are scalar, Bob's actual measurement in the $W$-basis ($W=X,Z$) is described by the filtering operation $F_w$ in Eqs.~\eqref{eqn-filtering1}-\eqref{eqn-filtering-Fx} (which is a diagonal matrix in the $W$-basis
and represents the efficiency mismatch) followed by the standard $W$-basis
measurement. 
Now, the artificial equalization can be described by another filter $\hat{F}_w$ which is diagonal in the $W$-basis such that 
$F_w \hat{F}_w =c I$ 
where $c$ is a constant independent of the basis $W$, meaning that data discarding removes the efficiency mismatch in both bases while incurring some loss.
Essentially, this is the same situation as Shor-Preskill's with a basis-independent loss,  
and thus their key rate formula can be applied.
Taking into account 
of the loss, the key generation rate for the data-discarding argument is
\begin{equation}
\label{eqn-example-scalar-noisy-discarding}
R_\text{noisy, scalar, discarding}
=
\frac{2 \min( \eta_0, \eta_1 ) }{\eta_0 + \eta_1} (1-H_2(e_p)- H_2(e_b)) .
\end{equation}
Here, $e_b$ and $e_p$ are the error rates after the data-discarding process corresponding to the $Z$-basis and $X$-basis measurement outcomes, 
respectively.
Also, the standard squash model for BB84 \cite{Gottesman2004,Tsurumaru2008,Beaudry2008} can be applied here to handle threshold detectors so that the incoming signals can always be treated as qubits and dark counts as random qubits.
This data-discarding argument in conjunction with the Shor-Preskill's proof bears similarity with our formulation.
In fact, the equalization filter $\hat{F}_x$ for the $X$-basis is the same as the virtual filter $G_{BT}$ in \myeqnref{eqn-BobVirtualFilter} in our formulation.
In our formulation, we assume that this virtual filter $G_{BT}$ is not and/or cannot be performed in practice and therefore we estimate the virtual phase error rate $e_p$ from the actual phase error rate $e_p'$ through problem \eqref{eqn-noisy-optimize-ep}.
On the other hand, with the data-discarding argument, this virtual filter is {\em actually} performed 
through the data-discarding process.
Thus, one can actually measure the ``virtual'' phase error rate $e_p$ directly 
(cf. Fig.~\ref{fig:measurement1-middle}) without solving problem \eqref{eqn-noisy-optimize-ep}.
This means that our estimation of the virtual phase error rate $e_p$ may be sub-optimal, and, in contrast, the data-discarding protocol allows us to obtain 
it
directly and accurately. Therefore, the data-discarding protocol
may give us a higher key generation rate than our general method.
A similar argument goes for the 
successful filtering probability $p_\text{succ,noisy}$.
This quantity can also be obtained without solving problem \eqref{eqn-noisy-optimize-psucc} since the virtual filter $G_{BT}$ is actually performed.
On the other hand, an analogous argument does not apply to the bit error rate since, in our formulation, no filtering is applied to the states to be measured in the $Z$ basis, whereas, in the data-discarding argument, the 
data-discarding filter $\hat{F}_z$ is applied.
In summary, the difference between our formulation and the data-discarding argument is that we assume that the virtual filter is not performed in practice.
By actually performing the filter, the data-discarding argument avoids solving
problems \eqref{eqn-noisy-optimize-psucc} and \eqref{eqn-noisy-optimize-ep}.
Nevertheless, one may proceed to solve problems \eqref{eqn-noisy-optimize-psucc} and \eqref{eqn-noisy-optimize-ep} with an addition constraint that
Eve's attack is
symmetric between bits ``0'' and ``1'' in both bases 
to obtain
$e_p=e_p'$ and $p_\text{succ,noisy}=\frac{2 \min( \eta_0, \eta_1 ) }{\eta_0 + \eta_1}$.
Using these results, our formulation in \myeqnref{eqn-noisy-final-key-rate} gives the key generation rate
\begin{equation}
\label{eqn-example-scalar-noisy-solvingP1P2}
R_\text{noisy, scalar}
=
\frac{2 \min( \eta_0, \eta_1 ) }{\eta_0 + \eta_1} (1-H_2(e_p))- H_2(e_b) .
\end{equation}
Note that the bit error rate $e_b$ here corresponds to that before data discarding.
By comparing 
Eqs.~\eqref{eqn-example-scalar-noisy-discarding} and \eqref{eqn-example-scalar-noisy-solvingP1P2}, we can see that the data-discarding argument gives a higher key generation rate than our general technique when the bit error rate remains the same before and after data discarding.
Nevertheless, this simple data-discarding argument does not extend to the general case with non-scalar efficiency matrices, 
whereas our general technique can handle this general case.

\subsection{Example 2: two-dimensional efficiency matrices}
\noindent
In this example, we illustrate the application of our proof to the non-trivial case with non-scalar efficiency matrices. 
Suppose that we have a polarization-coding QKD system in which the two detectors have imperfect efficiency responses to the arrival times of signals (assuming that only two arrival times are allowed for simplicity).
The two detectors have the 
following detection efficiency matrices:
\begin{eqnarray}
F_0^\dagger F_0 =
\begin{bmatrix}
.8 & -.2 \\
-.2 & .4
\end{bmatrix},
\phantom{xxxx}
F_1^\dagger F_1 =
\begin{bmatrix}
.3 & .1 \\
.1 & .9
\end{bmatrix} .
\end{eqnarray}
Here, the basis used to represent these matrices is the arrival times.
Thus, for example, the first detector responds with $80\%$ efficiency when hit by a signal arriving at the first time 
instant.
Also, the fact that the matrices are non-diagonal means that the detectors have non-trivial efficiency responses to signals entangled across the two time instants.

\begin{figure}
\subfigure[Probability of successful filtering.]{
\label{fig:keyrate-psucss}
\centerline{
\includegraphics[width=.5\columnwidth]{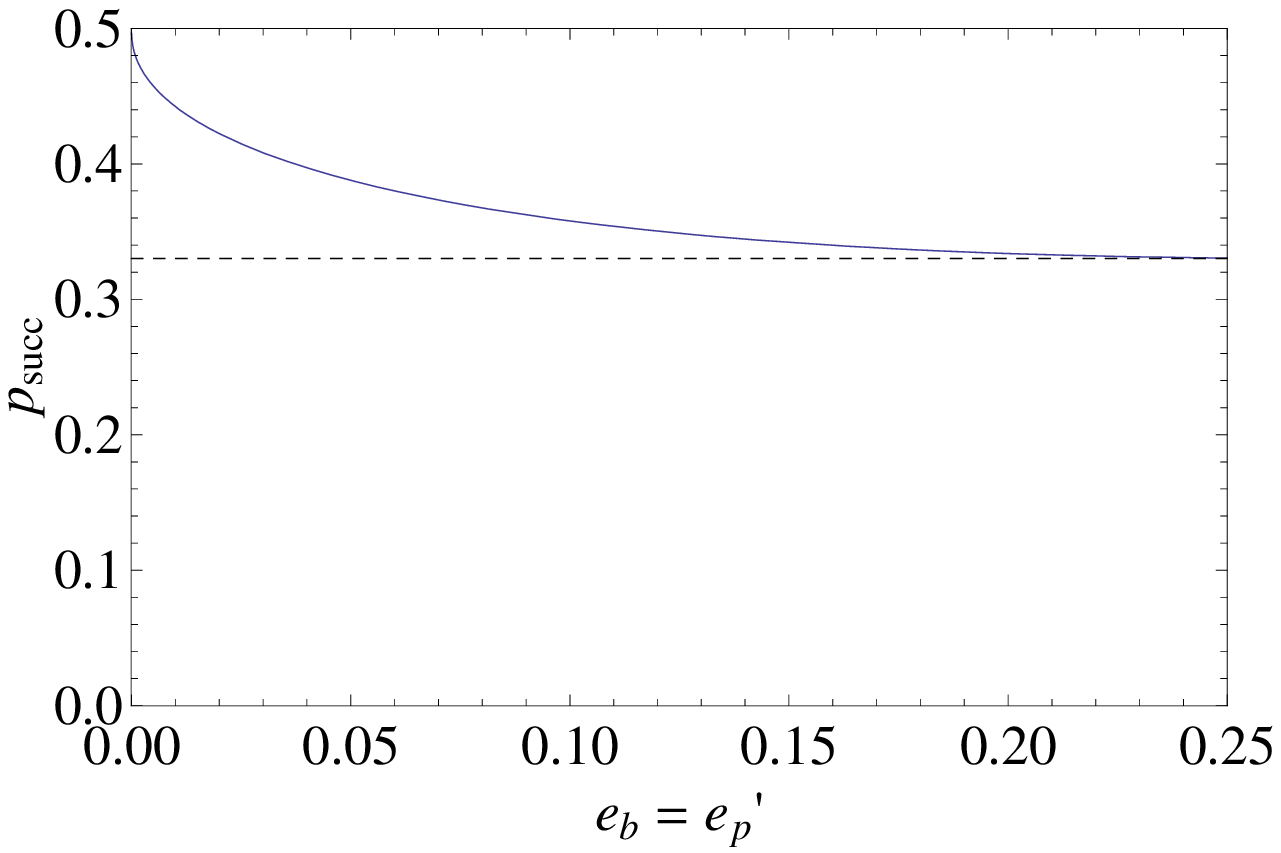}}
}
\\
\subfigure[Phase error rate.]{
\label{fig:keyrate-ep}
\centerline{
\includegraphics[width=.5\columnwidth]{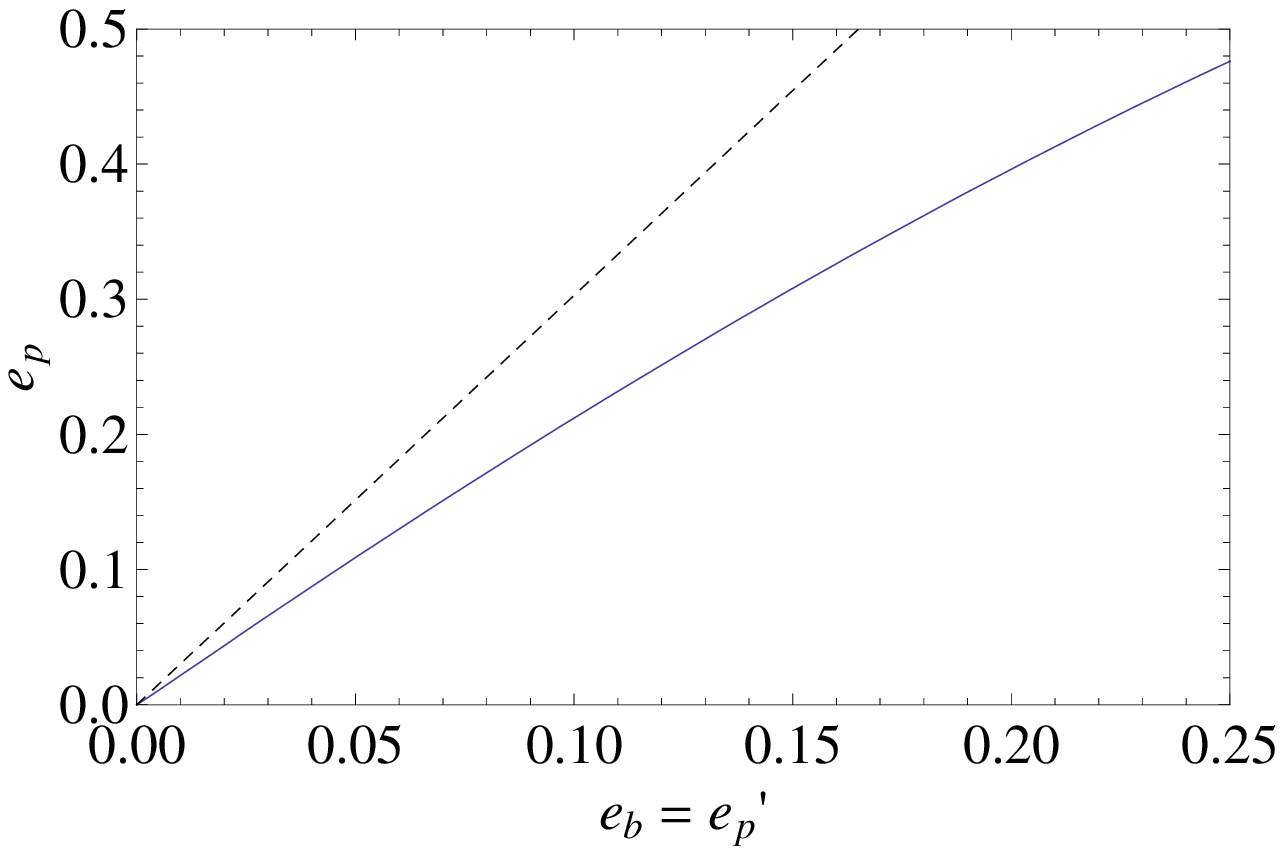}}
}
\\
\subfigure[Key generation rate.]{
\centerline{
\includegraphics[width=.5\columnwidth]{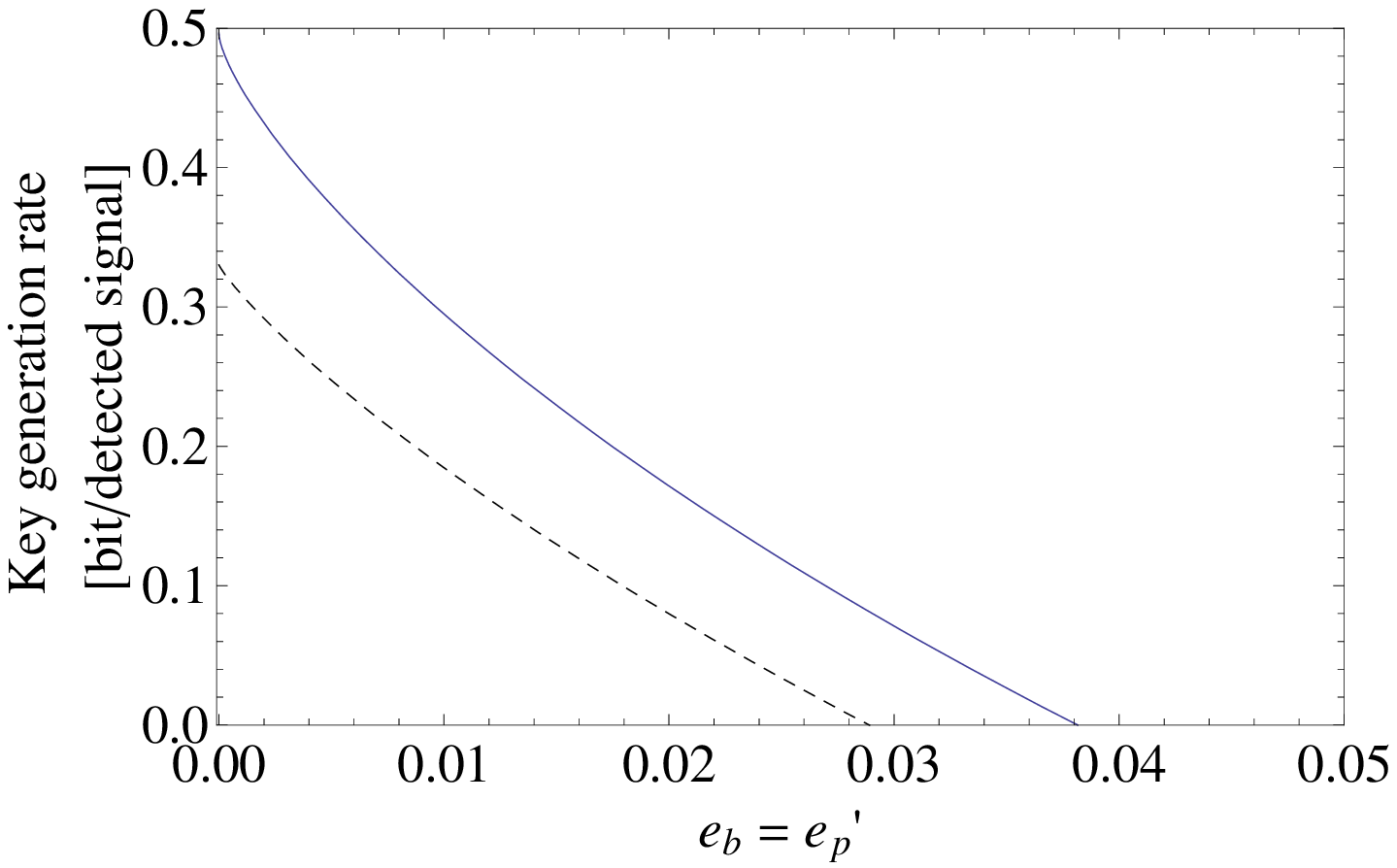}}
}
\fcaption{\label{fig:keyrate}
Efficiency mismatch example for two-dimensional efficiency matrices.  Solid curves are obtained from solving problems \eqref{eqn-noisy-optimize-psucc} and \eqref{eqn-noisy-optimize-ep}.  Dashed curves are obtained from 
the suboptimal bounds in Eqs.~\eqref{eqn-psucc-suboptimal1} and \eqref{eqn-ep-suboptimal1}.
The key generation rate is computed using \myeqnref{eqn-noisy-final-key-rate}.
The maximum efficiency mismatch ratio is $D_1=3.03$.
}
\end{figure}

We first compute the efficiency ratios between the two detectors, based on the above two matrices.
These ratios are the diagonal elements of the diagonal matrix $D$ which appear in the key generation rate expressions of \myeqnref{eqn-noiseless2} and \myeqnref{eqn-noisy-final-key-rate}. 
The ratios
are
the eigenvalues of $F_0^\dagger F_0 (F_1^\dagger F_1)^{-1}$ and are computed to be
\begin{eqnarray}
D_1 = 3.03 , \phantom{xxxx} D_2 = 0.356 .
\end{eqnarray}
To get the key generation rate in the noiseless case, we substitute $D$ into \myeqnref{eqn-noiseless2} and get
\begin{eqnarray}
R_\text{noiseless} &=& 0.496 .
\end{eqnarray}
To get the key generation rate for the general noisy case, we first compute the $C$ matrix from \myeqnref{eqn-matrixC}:
\begin{align}
C=
\begin{bmatrix}
0.51 & -0.17\\
0.12 & 0.56
\end{bmatrix}.
\end{align}
We then use $C$ in the expressions for the various probabilities in
Eqs.~\eqref{eqn-noisy-eb2}-\eqref{eqn-noisy-psucc2} and numerically solve for 
$p_\text{succ,noisy}$ and $e_p$ in problems \eqref{eqn-noisy-optimize-psucc} and \eqref{eqn-noisy-optimize-ep} for some given observed error rates.
Alternatively, one may determine suboptimal values of $p_\text{succ,noisy}$ and $e_p$ using Eqs.~\eqref{eqn-psucc-suboptimal1} and \eqref{eqn-ep-suboptimal1} without computing $C$.
Finally, the key generation rate is computed using
\myeqnref{eqn-noisy-final-key-rate}.
The key generation rate for the noisy case with the assumption of $e_b=e_p'$
is plotted in Fig.~\ref{fig:keyrate} along with the corresponding filtering probability and phase error rate.
Note that the minimum efficiency ratio is $1/D_1=0.330$ and the maximum is $D_1=3.03$.
These are shown as the dashed curve of Fig.~\ref{fig:keyrate-psucss} and the slope of the dashed curve of Fig.~\ref{fig:keyrate-ep}, respectively.
Even though the mismatch ratio is quite high, positive key generation rate can still be obtained.

\section{Detection scheme with four phase settings\label{sec-fourphase}}
\noindent
\begin{figure}
\includegraphics[width=1\columnwidth]{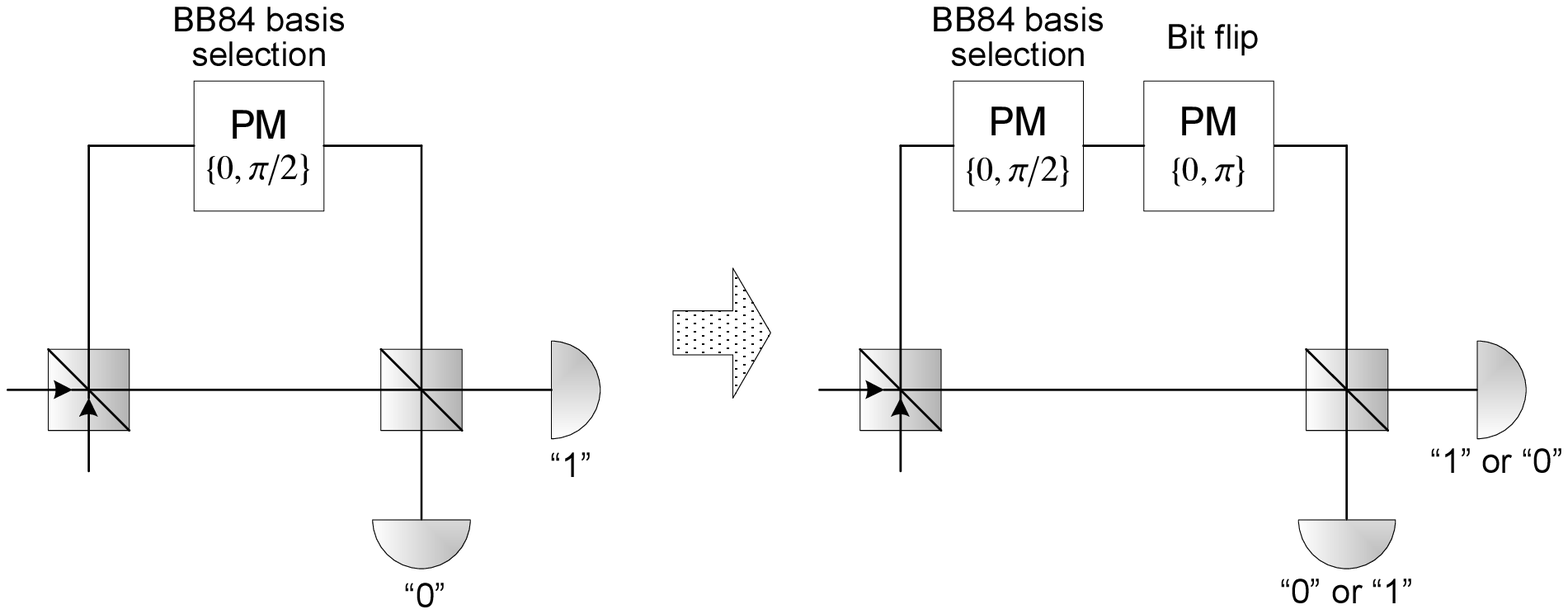}
\fcaption{\label{fig:fourphase}
The diagram on the left is the standard implementation of phase-encoding BB84 with a Mach-Zehner interferometer at Bob's side.
Bob selects the measurement basis with a phase modulator set to either $0$ or $\pi/2$.
The diagram on the right is the four-phase scheme where Bob randomly flips the bit value assignment of the two detectors with a phase modulator.
The two logical phase modulators can be implemented as one with four possible phases.
}
\end{figure}%
In this section, we prove 
how randomly switching the bit assignments of the two detectors for each quantum signal
can completely eliminate the effect of detection efficiency mismatch.
Previously, a four-phase modulation scheme was proposed \cite{Nielsen2001,LaGasse2005} to implement the BB84 protocol with only
one detector at Bob's side.
This scheme uses only one detector to measure both bit ``0'' and bit ``1'' .
Essentially, an additional bit flip operation is performed at random to prepare the detector for detecting bit ``0'' or bit ``1''.
So, half of the time, the detector is used for detecting bit ``0'' and the other half bit ``1''.
To implement this method in a phase-coding BB84 QKD system, Bob's phase modulator applies four phase settings $\{0,\pi/2,\pi,3\pi/2\}$ instead of the two $\{0,\pi/2\}$ in the normal case.
The motivation of the original proposals \cite{Nielsen2001,LaGasse2005} for this scheme was to save one detector but not for security purpose.
In fact,
since only one detector is used in this scheme, there is no issue of efficiency mismatch between detectors;
however, half the bits are lost, lowering the overall efficiency.
Moreover, we first
pointed out \cite{Qi2007a} that the
efficiency loophole could be closed by incorporating the four phase settings
in BB84 even with two detectors.
Here, we rigorously prove this claim.
Furthermore, we remark that the four-phase scheme can also be used to solve the detection efficiency mismatch problem that arises due to dead-times. 
In the following, we consider using two detectors together with the four phase settings 
in BB84.
In this case, the overall efficiency is not affected because two detectors are used instead of one.
The four phase settings are applied at Bob's side to ``average out'' the effect of the efficiency responses of the two detectors (see Fig.~\ref{fig:fourphase}).
At first sight, the key generation rate of this scheme may appear to be the average of the rates when Bob performs the bit flip operation and when he does not.
This is because if one considers entanglement distillation, the total entanglement available should just be the sum of the entanglement of the two cases.
However, when we are concerned with key generations, Bob's action on whether to flip or not is not known to Eve and thus serves to disentangle Eve further (i.e., it acts as a shield \cite{Horodecki2005}).
This allows the overall key generation rate to be higher than the average of the two cases.
In what follows, we show that indeed this scheme is capable of completely removing the effect of efficiency mismatch and the key generation rate for the no-mismatch case is recovered.

To begin, let's consider that Bob has a quantum coin that determines whether to apply a bit flip operation (which is the same as switching the two detectors).
The filtering operations $F_z'$ and $F_x'$ associated with the detectors' responses are (see Fig.~\ref{fig:measurement1})
\begin{eqnarray}
F_z' &=& 
F_z \otimes \ket{0}_C\bra{0} + X_B F_z X_B \otimes \ket{1}_C\bra{1}
\\
F_x' &=& 
F_x \otimes \ket{0}_C\bra{0} + Z_B F_x Z_B \otimes \ket{1}_C\bra{1}
\end{eqnarray}
where system $C$ represents the quantum coin, $F_z$ is from \myeqnref{eqn-filtering1}, and $F_x$ is from \myeqnref{eqn-filtering-Fx}.
The entire state after Eve's operation is [cf. \myeqnref{eqn-noisy-state-after-Eve2}, which does not involve the quantum coin]
\begin{eqnarray}
\ket{\Psi_4^{(l)}}
&=&
\sum_{i,j} E_{B}^{(l,i,j)} 
(\ket{00}+\ket{11})_{AB} \otimes (\ket{0} +\ket{1})_C \otimes \ket{\gamma(l,i,j)}_{T}
\otimes \ket{i}_E ,
\label{eqn-fourphase-state-after-Eve1}
\end{eqnarray}
where $E_{B}^{(l,i,j)}$ is defined in Appendix~\ref{app-simplification}.
We design the virtual filter $G_{BT,\text{4-phase}}$ so that the resulting state 
$G_{BT,\text{4-phase}} F_z' \ket{\Psi_4^{(l)}}$
[
Fig.~\ref{fig:measurement1-middle}]
is the same as the state used to determine the actual phase error probability $ F_x' \ket{\Psi_4^{(l)}}$ 
[
Fig.~\ref{fig:measurement1-bottom}].
In this way, the virtual $X$-basis measurement and the actual $X$-basis measurement exhibit the same phase error probabilities.

The state from which the actual phase error rate is estimated is
the state in \myeqnref{eqn-fourphase-state-after-Eve1} filtered with $F_x'$ (see 
Fig.~\ref{fig:measurement1-bottom}):
\begin{eqnarray}
&F_x' \ket{\Psi_4^{(l)}} & \nonumber \\
\label{eqn-fourphase-state-after-Fx}
=&
\sum_{i,j} 
\Big\{ &
\Big[
\frac{a_I+a_X}{\sqrt{2}} \ket{++}_{AB} +
\frac{a_Z-a_Y}{\sqrt{2}} \ket{-+}_{AB}
\Big]
\otimes 
( \ket{0}_C F_0\ket{\gamma}_{T} +\ket{1}_C F_1\ket{\gamma}_{T} ) \\
&+&
\Big[
\frac{a_I-a_X}{\sqrt{2}} \ket{--}_{AB} +
\frac{a_Z+a_Y}{\sqrt{2}} \ket{+-}_{AB}
\Big]
\otimes 
( \ket{0}_C F_1\ket{\gamma}_{T} +\ket{1}_C F_0\ket{\gamma}_{T} )  \Big\}
\otimes \ket{i}_E \nonumber
\end{eqnarray}
where $\ket{\gamma}_{T}=\ket{\gamma(l,i,j)}_{T}$ and $a_W=a_W^{(l,i,j)}, W=\{I,X,Y,Z\}$ for simplified notation.
Now, we design $G_{BT,\text{4-phase}}$ so that $G_{BT,\text{4-phase}} F_z' \ket{\Psi_4^{(l)}}$ is the same as \myeqnref{eqn-fourphase-state-after-Fx}.
Notice that
\begin{eqnarray}
&F_z' \ket{\Psi_4^{(l)}} & \nonumber \\
\label{eqn-fourphase-state-after-Fz}
=&
\sum_{i,j} 
\Big\{ &
\Big[
\frac{a_I+a_Z}{\sqrt{2}} \ket{00}_{AB} +
\frac{a_X+a_Y}{\sqrt{2}} \ket{10}_{AB}
\Big]
\otimes 
( \ket{0}_C F_0\ket{\gamma}_{T} +\ket{1}_C F_1\ket{\gamma}_{T} ) \\
&+&
\Big[
\frac{a_I-a_Z}{\sqrt{2}} \ket{11}_{AB} +
\frac{a_X-a_Y}{\sqrt{2}} \ket{01}_{AB}
\Big]
\otimes 
( \ket{0}_C F_1\ket{\gamma}_{T} +\ket{1}_C F_0\ket{\gamma}_{T} )  \Big\}
\otimes \ket{i}_E . \nonumber
\end{eqnarray}
We heuristically design $G_{BT,\text{4-phase}}=U_2 U_1$ in two steps.
First, we choose 
\begin{align}
U_1 =&\ket{0}_B \bra{0} \otimes I_C \otimes I_T +
\ket{1}_B \bra{1} \otimes X_C \otimes I_T
\end{align}
which is basically a CNOT operation in system $BC$.
This operation leads to
\begin{align}
&U_1 F_z' \ket{\Psi_4^{(l)}}  \nonumber \\
\label{eqn-fourphase-state-after-U1Fz}
=&
\sum_{i,j} E_{B}^{(l,i,j)} 
(\ket{00}+\ket{11})_{AB} \otimes \Big( \ket{0}_C \otimes F_0 \ket{\gamma(l,i,j)}_{T} + \ket{1}_C \otimes F_1 \ket{\gamma(l,i,j)}_{T} \Big)
\otimes \ket{i}_E .
\end{align}
Notice the similarity of \myeqnref{eqn-fourphase-state-after-U1Fz} to  \myeqnref{eqn-fourphase-state-after-Eve1} which represents the state before the detectors' responses are applied.
As a special case when there is no noise (i.e., $E_{B}^{(l,i,j)}=I$),  \myeqnref{eqn-fourphase-state-after-U1Fz} becomes the perfect EPR pair in system $AB$ tensor with system $CTE$, meaning that applying the virtual filter $U_1$ is already sufficient to completely eliminate the effect of detection efficiency mismatch.
In this special case, the probability of successful virtual filtering is $1$ (since $U_1$ is unitary) and thus the key generation rate on detected signals is $1$.

Let's continue to design $U_2$, which is needed for the noisy case.
By choosing
\begin{align}
U_2 =&\ket{+}_B \bra{+} \otimes I_C \otimes I_T +
\ket{-}_B \bra{-} \otimes X_C \otimes I_T ,
\end{align}
one can easily verify that $U_2 U_1 F_z' \ket{\Psi_4^{(l)}} = F_x' \ket{\Psi_4^{(l)}}$.
Thus, the final virtual filter is $G_{BT,\text{4-phase}}=U_2 U_1$.
Note that $G_{BT,\text{4-phase}}$ is not $F_x' F_z'^{-1}$ in general 
(i.e., we did not invert the filters in an naive way).
To find the final key generation rate on detected signals, we may use  
\myeqnref{eqn-noisy-final-key-rate}
with the probability of successful virtual filtering $p_\text{succ,noisy}=1$ (since $G_{BT,\text{4-phase}}$ is unitary) to get
\begin{eqnarray}
\label{eqn-fourphase-key-rate1}
R_\text{4-phase} &=&
1-H_2(e_p) - H_2(e_b)
\end{eqnarray}
where $e_p$ is the phase error rate estimated using the actual $X$-basis measurement (see 
Fig.~\ref{fig:measurement1-bottom}) and $e_b$ is the bit error rate estimated using the actual $Z$-basis measurement (see Fig.~\ref{fig:measurement1-top}).
Apparently, this key generation rate in \myeqnref{eqn-fourphase-key-rate1} is the same as if there is no efficiency mismatch (e.g., Refs.~\cite{Shor2000,Lo2005b}).
This means that the effect of the efficiency mismatch is completely removed by using this four-state scheme.

\section{Multi-photon signals\label{sec-multi-photon}}
\noindent
In practice, the channel may receive multi-photon signals from Alice and may emit multi-photon signals to Bob.
We discuss this issue here.

\subsection{Input with multi-photon signals\label{sec-multi-photon-input}}
\noindent
Since our detector model and proof work on the assumption that the input to Bob are single-photon signals, applying our proof to practice settings where multi-photon signals may be present requires special attention.
In order to cope with this practical issue, one may consider implementing the detector-decoy idea of Ref.~\cite{Moroder2008} to estimate the fraction of single-photon input to Bob.
The detector-decoy idea involves the receiver Bob adding
an attenuator in front of his detector to monitor the transmission
properties
(e.g., transmittance and quantum bit error rate) as a function of
attenuation.
By solving linear equations related to these properties,
one can infer the fraction of single-photon input to Bob.
Signals received by Bob are randomly chosen to be used for this estimation process or the normal key generation process.
Upon knowing the fraction of single-photon input signals, one can apply the tagged-signals idea of \cite{Gottesman2004} to compute the overall key generation rate by assuming that all multi-photon input signals are insecure and only single-photon input signals are distillable.
Our proof provides a way to compute the number of secure bits in
the single-photon parts, and this is what is needed when we 
apply \cite{Gottesman2004} 
to extend the situation with multi-photon input signals.
When applying our proof in this case, one may also assume pessimistically that all errors come from the single-photon parts.
Note that when applying the idea of \cite{Gottesman2004}, we only need to know the fraction of the single-photon input signals but not their positions.

\subsection{Phase-randomized weak coherent source}
\noindent
Our proof can also be applied to the case with a phase-randomized weak coherent source, with or without decoy states \cite{Hwang2003,Lo2005,Ma2005b,Wang2005a,Wang2005b,Harrington2005,Zhao2006,Zhao2006b},
by following the argument in \cite{Gottesman2004}.
Essentially, 
to incorporate 
both multi-photon inputs and outputs,
one estimates the fraction of single-photon input signals that originated from single-photon outputs, and applies the result of \cite{Gottesman2004}.

\section{Concluding remark\label{sec-conclusion}}
\noindent
In practice, it is hard to build two identical detectors.
Owing to widespread existence of detector dead-times or different detector responses
as functions of some auxiliary variables in, for example, time, frequency or spatial domain,
two detectors almost certainly exhibit detection efficiency mismatch.
For practical QKD systems to be secure, it is thus important to
prove the security of QKD systems with detection efficiency mismatch.
In this paper, we prove the security of the BB84 protocol when the detectors respond to some auxiliary domain, in addition to the qubit space representing the information bit in the normal case.
Specifically, we show that once the detectors' responses to the auxiliary domain ($F_i$) are characterized, we can obtain an amount of privacy amplification sufficient to remove Eve's information on the final key.
We show that this amount is directly related to the 
maximum efficiency ratio between the two detectors.

We show that the detectors' responses $F_i$ can be characterized with a finite number of samples even when time is the auxiliary domain by using a narrow-band frequency filter.
Thus, we may test $F_i$ with a finite number of test signals in practice in order to characterize it. 
One issue about the applicability of our proof is that we assume that the detectors' responses $F_i$ are stable over time.
Once we have stable estimates of $F_i$ from testing the detectors, our proof can be applied to obtain the final key generation rate.

The key generation rate derived in this paper may not be optimal and there may be ways to improve it.
Our speculation is based on 
two observations: one in the noiseless case and another in the noisy case.
For the first observation in the noiseless case,
suppose that the two detectors have different scalar efficiencies, i.e. $F_i=\eta_i$.
In this case, the state shared by Alice and Bob is $\sqrt{\eta_0} \ket{00} + \sqrt{\eta_1} \ket{11}$.
Since there are $2^{N h_2(\eta_0/(\eta_0+\eta_1))}$ typical strings, the key generation rate is simply $h_2(\eta_0/(\eta_0+\eta_1))$, which may be obtained by applying the appropriate amount of privacy amplification (see also \cite{Bennett1996}).
On the other hand, our current proof yields a key generation rate of $2\min(\eta_0,\eta_1)/(\eta_0+\eta_1)$ (see \myeqnref{eqn-noiseless3}), which is in general smaller.
For the second observation in the noisy case,
we see in Example~1 (see Sec.~\ref{subsec-example1}) that a simple data-discarding argument can produce a higher key generation rate than our proof (when having scalar efficiencies).
Thus, we speculate that there may be ways to improve the key generation rate even in the general noisy case where non-scalar detection efficiencies $F_i$ are used.
The source of sub-optimality of our current proof may be that we distinguish non-orthogonal states on Bob's side on each signal independently.
Thus, it would be interesting to apply a
collective method for entanglement concentration, instead of
the Procrustean method for entanglement concentration, to the context of
QKD.
We leave this potential improvement for future work.

We have shown rigorously that using four phase settings in BB84 can equalize the
detection efficiencies for bits ``0'' and ``1'', thus solving the detection efficiency mismatch problem.
However, as noted in our introduction,
Eve may try to break this four-state-measurement counter-measure
by using a combined strong pulse attack and time-shift attack
\cite{Lo2007Tropical,Lydersen2008}.
This demonstrates that counter-measures may lead to new attacks.

\nonumsection{Acknowledgments}
\noindent
We thank 
J.-C. Boileau and D. Gottesman
for enlightening discussions. Support of the
funding agencies CFI, CIPI, the CRC program, CIFAR, MITACS, NSERC,
QuantumWorks, OIT, and PREA is gratefully acknowledged. 
C.-H. F. Fung 
gratefully
acknowledges 
support from the Postdoctoral Fellowship program of NSERC of Canada and the RGC
grant No. HKU~701007P of the HKSAR Government.

\nonumsection{References}
\bibliographystyle{IEEEtran2}
\bibliography{paperdb}

\appendix{: Analysis of time-dependent efficiency\label{app-time-finite}}
\noindent
Here, we justify that the efficiency response of a detector can be characterized with a finite number of samples when a narrow-band filter is placed before the detector.

Before we begin, we remark that 
we will work on the baseband signals,
even though the signals are modulated to a higher carrier frequency before detection.
This is valid because practical detectors 
are not fast enough to respond to the optical frequency of the input signal.
So only the envelope of the signal, which is equivalent to the baseband of the signal, will be detected.
Thus, with this built-in demodulation function in the detectors, 
we can work on the baseband signals.

Suppose that the input (quantum) signal at the detector location is $E(t)$, and the Gaussian-shaped frequency filter is $g(f)$ and has an effective bandwidth of $B$.
Since the filter has a fixed bandwidth, we can assume without loss of generality that the input signal also has a fixed bandwidth $B$.
Then the input signal in the frequency domain can be expressed as
\begin{align}
E(f) &= [E(f) * \sum_{k=-\infty}^\infty \delta(f - 2 B k)] h(f)
\end{align}
where $h(f)$ is the perfect rectangular filter with bandwidth $B$, and the asterisk denotes convolution.
Here, we are essentially repeating $E(f)$ indefinitely with a separation of $2B$ and then chopping the spectrum with the low-pass filter $h(f)$.
This can be seen by noting that 
$E(f) * \delta(f-2B) = E(f-2B)$.
This step may look superfluous, but the reason for this will be clear when we look at the time domain.
The output of the filter is, in the frequency domain, 
\begin{align}
E'(f) &= E(f) g(f) \\
&= [E(f) * \sum_{k=-\infty}^\infty \delta(f - 2 B k)] g(f)
\end{align}
where we have used the fact that $g(f)=g(f) h(f)$.
In the time domain, the output is obtained through the Fourier transform:
\begin{align}
E'(t) &= (2B)^{-1} [E(t) \sum_{k=-\infty}^\infty \delta(t-k/(2B))] * g(t) . \label{eqn-detector-model-filter-output1}
\end{align}
Here, we have used the convolution property of the Fourier transform which says that convolution in the time domain becomes multiplication in the frequency domain and vice versa.
Also, we used the fact that the Fourier transform of an impulse train is also an impulse train, i.e.,
\begin{align}
\sum_{k=-\infty}^\infty \delta(f - 2 B k) \leftrightarrow
(2B)^{-1}\sum_{k=-\infty}^\infty \delta(t-k/(2B))
\end{align}
are a Fourier transform pair.
Finally, we expand the convolution in Eq.~\eqref{eqn-detector-model-filter-output1} to get
\begin{align}
E'(t) &= (2B)^{-1} \sum_{k=-\infty}^\infty E(k/(2B)) g(t-k/(2B)) .
\end{align}

This last equation illustrates the key point, which is that given any input signal $E(t)$, the output of the filter can always be represented by
a train of Gaussian pulses (possibly with different amplitudes) spaced at $1/(2B)$ time intervals.
Thus, the detector always receives a train of Gaussian pulses located at fixed time instants and if the fixed length of the gating window is taken into account, one can characterize the detector with a finite number of pulses sent within the gating window.
Therefore, the detection efficiency response ${F_i}^\dagger F_i$ effectively 
becomes finite dimensional.
This proves that with an addition of a frequency filter, the characterization of the time-dependent efficiency of the detectors becomes discrete and finite-dimensional, and thus is readily applicable to our security proof.
Note that although our analysis does not explicitly involve quantum mechanics, the analysis is still applicable to quantum states since the Fourier transform is an unitary transformation (also linear) that changes between the time basis and the frequency basis, and the quantities $E(f)$ and $E(t)$ can be regarded as the amplitudes of a quantum state in the respective basis.

\appendix{: Determination of Bob's virtual filter for the noiseless case\label{app-virtual-filter}}
\noindent
We determine Bob's virtual filter $G_{BT}$ in Eq.~\eqref{eqn-BobVirtualFilter} by finding the matrix $C \in \field{C}^{d\times d}$.
Since a valid filter $G_{BT}$ must satisfy $G_{BT}^\dagger G_{BT} \leq I$, we find $C$ such that this is satisfied while 
maximizing the key generation rate in Eq.~\eqref{eqn-noiseless1}.
This problem can be expressed as
\begin{eqnarray}
\label{eqn-maxminprob1}
\max_{C} \min_{\ket{\gamma}} && \frac{2\bra{\gamma} (C^\dagger C)_T \otimes I_E \ket{\gamma}_{TE}}{\bra{\gamma} (F_0^\dagger F_0 + F_1^\dagger F_1)_T \otimes I_E \ket{\gamma}_{TE}}  \\
s.t. && G_{BT}^\dagger G_{BT} \leq I .
\end{eqnarray}
Note that since we are interested in the worst-case key generation rate, we form a max-min problem as opposed to a min-max problem, in light of the max-min inequality: $\max_a \min_b $ $ f(a,b) \leq \min_b \max_a f(a,b)$.

To solve this optimization problem, first note that the condition $G_{BT}^\dagger G_{BT} \leq I$ is the same as the condition $G_{BT} G_{BT}^\dagger \leq I$, which can be expanded (by using Eq.~\eqref{eqn-BobVirtualFilter}) as
\begin{eqnarray}
\label{eqn-filterCondition1}
	\ket{0_z}_B \bra{0_z} \otimes C (F_0^\dagger F_0)^{-1} C^\dagger + \ket{1_z}_B \bra{1_z} \otimes  C (F_1^\dagger F_1)^{-1} C^\dagger	\leq I .
\end{eqnarray}
Letting $C=C_1 U^\dagger F_0$ where $C_1 \in \field{C}^{d\times d}$ will be determined next, $U \in \field{C}^{d\times d}$ is the unitary matrix of the Hermitian decomposition of $F_0 (F_1^\dagger F_1)^{-1} F_0^\dagger = U D U^\dagger$, and $D=\text{diag}(D_1,\ldots,D_d)$ is a diagonal matrix with positive real elements, \myeqnref{eqn-filterCondition1} can be expressed as
\begin{eqnarray}
\label{eqn-filterCondition2}
	\ket{0_z}_B \bra{0_z} \otimes C_1 C_1^\dagger + \ket{1_z}_B \bra{1_z} \otimes  C_1 D C_1^\dagger	\leq I.
\end{eqnarray}
This allows us to redefine the constraints of the problem.

Next, we consider the objective function in \myeqnref{eqn-maxminprob1}, which can be simplified as
\begin{eqnarray}
&& \frac{2\bra{\gamma} (C^\dagger C)_T \ket{\gamma}}{\bra{\gamma} (F_0^\dagger F_0 + F_1^\dagger F_1)_T \ket{\gamma}} \\
&=&
\frac{2\bra{\gamma'} ({F_0^\dagger}^{-1} C^\dagger C F_0^{-1})_T \ket{\gamma'}}{\bra{\gamma'} (I +  {F_0^\dagger}^{-1} F_1^\dagger F_1 F_0^{-1})_T \ket{\gamma'}} \\
&=&
\frac{2\bra{\gamma''} (C_1^\dagger C_1 )_T \ket{\gamma''}}{\bra{\gamma''} (I +  D^{-1})_T \ket{\gamma''}} \\
&=&
\frac{2\bra{\gamma'''} ((I +  D^{-1})^{-1/2} C_1^\dagger C_1 (I +  D^{-1})^{-1/2})_T \ket{\gamma'''}}{\langle{\gamma'''} \ket{\gamma'''}} .
\end{eqnarray}
Now, the problem can be re-written as
\begin{eqnarray}
\label{eqn-maxminprob2}
R_\text{noiseless}=
\max_{C_2} \min_{\ket{\gamma}} && \frac{2\bra{\gamma} (C_2^\dagger C_2)_T \otimes I_E \ket{\gamma}_{TE}}{\langle{\gamma} \ket{\gamma}}  \\
s.t. && C_2 (I +  D^{-1})C_2^\dagger \leq I \\
&& C_2 (I +  D)C_2^\dagger \leq I
\end{eqnarray}
where $C_2=C_1 (I +  D^{-1})^{-1/2}$.
We further simply the problem as
\begin{eqnarray}
\label{eqn-maxminprob3}
R_\text{noiseless}=&&
2 \max_{C_2} 
[\text{minimum eigenvalue of }
C_2^\dagger C_2]\\
s.t. && C_2
\begin{bmatrix}
1+\frac{1}{D_1} \\
& \ddots \\
&& 1+\frac{1}{D_d}
\end{bmatrix}
C_2^\dagger \leq I \\
&& C_2 
\begin{bmatrix}
1+D_1 \\
& \ddots \\
&& 1+D_d
\end{bmatrix}
C_2^\dagger \leq I .
\end{eqnarray}
The solution is
\begin{eqnarray}
\label{eqn-matrixC2}
C_2 &=& 
\begin{bmatrix}
\sqrt{\min\left(\frac{1}{1+D_1},\frac{D_1}{1+D_1}\right)} \\
& \ddots \\
&& \sqrt{\min\left(\frac{1}{1+D_d},\frac{D_d}{1+D_d}\right)}
\end{bmatrix}\\
\label{eqn-matrixC}
C &=& 
\begin{bmatrix}
\sqrt{\min\left(\frac{1}{D_1},1\right)} && \\
& \ddots & \\
&& \sqrt{\min\left(\frac{1}{D_d},1\right)}
\end{bmatrix} U^\dagger F_0, 
\end{eqnarray}
and the final key generation rate can be obtained 
by substituting \myeqnref{eqn-matrixC2} into \myeqnref{eqn-maxminprob3}:
\begin{align*}
R_\text{noiseless}
&=
2 \min \left(\frac{1}{1+D_1}, \frac{1}{1+1/D_1}, \frac{1}{1+D_2},\ldots\right) \\
&=
\frac{2}{1+\max\left(D_1,\frac{1}{D_1},\ldots,D_d,\frac{1}{D_d}\right)} .
\end{align*}

\appendix{: Simplification of Eqs.~(\ref{eqn-noisy-eb1})-(\ref{eqn-noisy-psucc1})\label{app-simplification}}
\noindent
Here, we show how to simplify $e_b$ in \myeqnref{eqn-noisy-eb1}.
The other quantities in Eqs.~\eqref{eqn-noisy-ep1}-\eqref{eqn-noisy-psucc1} can be similarly simplified.

Let's re-write \myeqnref{eqn-noisy-state-after-Eve} 
by separating the operations for systems $B$ and $T$ by expressing $E_{BT}^{(l,i)}=\sum_j E_B^{(l,i,j)} \otimes E_T^{(l,i,j)}$:
\begin{eqnarray}
\ket{\Psi_2^{(l)}} &=&
\sum_{i,j} E_{B}^{(l,i,j)} 
(\ket{00}+\ket{11})_{AB} \otimes \ket{\gamma^{(l,i,j)}}_{T}
\otimes \ket{i}_E
\label{eqn-noisy-state-after-Eve2}
\end{eqnarray}
where $\ket{\gamma^{(l,i,j)}}=E_{T}^{(l,i,j)}\ket{0}$.
Consider \myeqnref{eqn-noisy-pre-virtual-filtered-state1} which is expanded as
\begin{eqnarray}
F_z \ket{\Psi_2^{(l)}} 
&=&
\sum_{i,j} 
\Big\{
\Big[
a_{00}^{(l,i,j)} \ket{00}_{AB} +
a_{10}^{(l,i,j)} \ket{10}_{AB} \Big]
\otimes F_0 \ket{\gamma(l,i,j)}_{T} +\\
&&
\Big[
a_{01}^{(l,i,j)} \ket{01}_{AB} +
a_{11}^{(l,i,j)} \ket{11}_{AB}
\Big]
\otimes F_1  \ket{\gamma(l,i,j)}_{T}
\Big\}
\otimes \ket{i}_E \nonumber
\end{eqnarray}
where
\begin{align}
a_{00}^{(l,i,j)} &= \frac{a_I^{(l,i,j)}+a_Z^{(l,i,j)}}{\sqrt{2}} &
a_{01}^{(l,i,j)} &= \frac{a_X^{(l,i,j)}-a_Y^{(l,i,j)}}{\sqrt{2}} \\
a_{10}^{(l,i,j)} &= \frac{a_X^{(l,i,j)}+a_Y^{(l,i,j)}}{\sqrt{2}} &
a_{11}^{(l,i,j)} &= \frac{a_I^{(l,i,j)}-a_Z^{(l,i,j)}}{\sqrt{2}}
\end{align}
and
Eve's operation is expressed in terms of the Pauli matrices as $E_{B}^{(l,i,j)}=a_I^{(l,i,j)} I + a_X^{(l,i,j)} X + a_Y^{(l,i,j)} i Y + a_Z^{(l,i,j)} Z$.  
Using this notation in \myeqnref{eqn-noisy-eb1}, we have
\begin{align}
e_b
=&
\frac{
\sum_{l,i,j,j'}
{a_{10}^{J'}}^\dagger a_{10}^J \bra{\gamma^{J'}} F_0^\dagger F_0  \ket{\gamma^J} +
{a_{01}^{J'}}^\dagger a_{01}^J \bra{\gamma^{J'}} F_1^\dagger F_1  \ket{\gamma^J}
}{
\sum_{l,i,j,j'}
{(a_{10}^{J'}}^\dagger a_{10}^J + {a_{00}^{J'}}^\dagger a_{00}^J) \bra{\gamma^{J'}} F_0^\dagger F_0  \ket{\gamma^J} +
{(a_{01}^{J'}}^\dagger a_{01}^J + {a_{11}^{J'}}^\dagger a_{11}^J) \bra{\gamma^{J'}} F_1^\dagger F_1  \ket{\gamma^J}
}
\end{align}
where, for simplicity,
$J'$ means $(l,i,j')$ and
$J$ means $(l,i,j)$.
Focusing on the first term in the numerator with fixed $l$ and $i$,
it is equal to
\begin{align}
& \sum_{j,j'}
{a_{10}^{J'}}^\dagger a_{10}^{J} \bra{\gamma^{J'}} F_0^\dagger F_0  \ket{\gamma^J} \\
=&
\begin{pmatrix}
\bra{\gamma^{(l,i,0)}} & \bra{\gamma^{(l,i,1)}} & \cdots
\end{pmatrix}
\left[
\begin{pmatrix}
a_{10}^{(l,i,0)\dagger} a_{10}^{(l,i,0)} & a_{10}^{(l,i,0)\dagger} a_{10}^{(l,i,1)} & \cdots \\
a_{10}^{(l,i,1)\dagger} a_{10}^{(l,i,0)} & \ddots \\ 
\vdots
\end{pmatrix}
\otimes
F_0^\dagger F_0
\right]
\begin{pmatrix}
\ket{\gamma^{(l,i,0)}}\\
\ket{\gamma^{(l,i,1)}}\\
\vdots
\end{pmatrix}.
\end{align}
The second matrix can immediately be recognized as $A^{(l,i)\dagger} \tilde{Z}_{10} A^{(l,i)}$
where $\tilde{Z}_{10}$ is from \myeqnref{eqn-4x4-matrices} below and
\begin{align}
A^{(l,i)} &= 
\begin{pmatrix}
a_I^{(l,i,0)} & a_I^{(l,i,1)} & \cdots \\
a_X^{(l,i,0)} & a_X^{(l,i,1)} & \cdots \\
a_Y^{(l,i,0)} & a_Y^{(l,i,1)} & \cdots \\
a_Z^{(l,i,0)} & a_Z^{(l,i,1)} & \cdots \\
\end{pmatrix}	.
\end{align}
By letting 
\begin{align}
\ket{\phi(l,i)}=(A^{(l,i)} \otimes I) \begin{pmatrix}
\ket{\gamma^{(l,i,0)}}\\
\ket{\gamma^{(l,i,1)}}\\
\vdots
\end{pmatrix}
\end{align}
where the identity matrix has dimensions $d \times d$
and applying similar arguments to other terms,
we have
\begin{align}
e_b
&=
\frac
{
\sum_{l,i}
\bra{\phi(l,i)} [\tilde{Z}_{10} \otimes F_0^\dagger F_0 +  \tilde{Z}_{01} \otimes F_1^\dagger F_1 ] \ket{\phi(l,i)}
}
{
\sum_{l,i}
\bra{\phi(l,i)} [(\tilde{Z}_{00}+\tilde{Z}_{10}) \otimes F_0^\dagger F_0 +  (\tilde{Z}_{11}+\tilde{Z}_{01}) \otimes F_1^\dagger F_1 ] \ket{\phi(l,i)}
}
\label{eqn-app-simplify-eb-1}
\end{align}
where $\tilde{Z}_{i,j}$ are constant matrices shown below in \myeqnref{eqn-4x4-matrices}.
Finally, by letting $\rho_E=\sum_{l,i} \ket{\phi(l,i)}
\bra{\phi(l,i)}$, \myeqnref{eqn-app-simplify-eb-1} becomes \myeqnref{eqn-noisy-eb2}.
Similarly, Eqs.~\eqref{eqn-noisy-ep1}-\eqref{eqn-noisy-psucc1} can be simplified where
$\tilde{Z}_{i,j}$ and $\tilde{X}_{i,j}$ are
\begin{align}
\tilde{Z}_{00} &= P( [1, 0, 0, 1]^\dagger )/2
&
\tilde{X}_{++} &= P( [1, 1, 0, 0]^\dagger )/2 \nonumber \\
\tilde{Z}_{10} &= P( [0,1,1,0]^\dagger )/2
&
\tilde{X}_{-+} &= P( [0, 0,-1, 1 ]^\dagger )/2 \nonumber \\
\tilde{Z}_{01} &= P( [0,1,-1,0]^\dagger )/2
&
\tilde{X}_{+-} &= P( [0, 0,1, 1 ]^\dagger )/2 \label{eqn-4x4-matrices} \\
\tilde{Z}_{11} &= P( [1, 0, 0, -1]^\dagger )/2
&
\tilde{X}_{--} &= P( [1, -1, 0, 0]^\dagger )/2 . \nonumber
\end{align}
Here, $P(\ket{\cdot})=\ket{\cdot} \bra{\cdot}$ is the projection operator.

\appendix{: Solving for the suboptimal bounds 
 \label{app-solve-suboptimal}}
\noindent
Here, we prove Eqs.~\eqref{eqn-psucc-suboptimal1} and \eqref{eqn-ep-suboptimal1}.
First, we consider solving
\begin{align}
p_\text{succ,noisy}
& \geq \min_{\ket{\phi}}
\frac
{
\bra{\phi} I \otimes C^\dagger C \ket{\phi}
}
{
\bra{\phi} [(\tilde{Z}_{00}+\tilde{Z}_{10}) \otimes F_0^\dagger F_0 +  (\tilde{Z}_{11}+\tilde{Z}_{01}) \otimes F_1^\dagger F_1 ] \ket{\phi}
}
\label{eqn-app-suboptimal-prove-psucc1}
\end{align}
where the right hand side comes from \myeqnref{eqn-noisy-psucc2} and
$\tilde{Z}_{00}+\tilde{Z}_{10}+\tilde{Z}_{11}+\tilde{Z}_{01}=I$.
Here, we only need to focus on rank-one $\rho_E=\ket{\phi}\bra{\phi}$ because of the following claim.
\noindent\begin{claim}
{\rm
\label{claim-ratio}
Given two ratios, $\frac{a_1}{a_2}$ and $\frac{b_1}{b_2}$, where $a_1,a_2,b_1,b_2 \in \field{R}_+$,
if $\frac{a_1}{a_2} \geq \frac{b_1}{b_2}$,
then $\frac{a_1}{a_2} \geq \frac{a_1+b_1}{a_2+b_2}$.
Similarly, 
if $\frac{a_1}{a_2} \leq \frac{b_1}{b_2}$,
then $\frac{a_1}{a_2} \leq \frac{a_1+b_1}{a_2+b_2}$.
}
\end{claim}%
This claim basically means that we only need to focus on the smallest or the largest ratio when there are more than one ratio to optimize over.
By substituting $\ket{\phi'}=[(\tilde{Z}_{00}+\tilde{Z}_{10}) \otimes F_0+(\tilde{Z}_{11}+\tilde{Z}_{01}) \otimes F_1] \ket{\phi}$ into the right hand side of \myeqnref{eqn-app-suboptimal-prove-psucc1}, 
we get
\begin{align}
p_\text{succ,noisy}
& \geq \min_{\ket{\phi}}
\frac
{
\bra{\phi'} [(\tilde{Z}_{00}+\tilde{Z}_{10}) \otimes F_0^{-\dagger} C^\dagger C F_0^{-1} +  (\tilde{Z}_{11}+\tilde{Z}_{01}) \otimes F_1^{-\dagger} C^\dagger C F_1^{-1} ] \ket{\phi'}
}{
\langle\phi' \ket{\phi'}
} .
\label{eqn-psucc-eigeneqn1}
\end{align}
By noting that $\tilde{Z}_{00}+\tilde{Z}_{10}$ and $\tilde{Z}_{11}+\tilde{Z}_{01}$ are orthogonal, the minimum eigenvalue of \myeqnref{eqn-psucc-eigeneqn1} is the minimum of the eigenvalues of $F_i^{-\dagger} C^\dagger C F_i^{-1}$ where $i=0,1$.
By using \myeqnref{eqn-matrixC}, one can immediately see that the 
eigenvalues of $F_0^{-\dagger} C^\dagger C F_0^{-1}$ are 
\begin{align}
\min\left(\frac{1}{D_i},1\right)  && i=1,\ldots,d
\label{eqn-eigenvalues-F0}
\end{align}
and that of $F_1^{-\dagger} C^\dagger C F_1^{-1}$ are
\begin{align}
\min(D_i,1) && i=1,\ldots,d .
\label{eqn-eigenvalues-F1}
\end{align}
Taking the minimum of these two sets of eigenvalues gives \myeqnref{eqn-psucc-suboptimal1}.

Next, we consider the maximization of $e_p/e_p'$ in \myeqnref{eqn-ep-suboptimal1}.
Using Eqs.~\eqref{eqn-noisy-ep2} and \eqref{eqn-noisy-epp2} (and Claim~\ref{claim-ratio}), we have
\begin{align}
\frac{e_p}{e_p'} = &
\frac{
\bra{\phi}\tilde{X}_{-+} \otimes C^\dagger C \ket{\phi} +
\bra{\phi}\tilde{X}_{+-} \otimes C^\dagger C \ket{\phi}
}{
\bra{\phi}\tilde{X}_{-+} \otimes F_0^\dagger F_0 \ket{\phi} +
\bra{\phi}\tilde{X}_{+-} \otimes F_1^\dagger F_1 \ket{\phi}
} \times \nonumber \\
&
\frac{
\bra{\phi}(\tilde{X}_{++}+\tilde{X}_{-+}) \otimes F_0^\dagger F_0 \ket{\phi} +
\bra{\phi}(\tilde{X}_{--}+\tilde{X}_{+-}) \otimes F_1^\dagger F_1 \ket{\phi}
}{
\bra{\phi}(\tilde{X}_{++}+\tilde{X}_{-+}) \otimes C^\dagger C \ket{\phi} +
\bra{\phi}(\tilde{X}_{--}+\tilde{X}_{+-}) \otimes C^\dagger C \ket{\phi}
} .
\label{eqn-ep-suboptimal2}
\end{align}
According to Claim~\ref{claim-ratio}, the first ratio in \myeqnref{eqn-ep-suboptimal2} is upper bounded by
the maximum of the eigenvalues of 
$F_i^{-\dagger} C^\dagger C F_i^{-1}$ where $i=0,1$; while the second ratio is upper bounded by
inverse of the minimum of the same set of eigenvalues.
Therefore, using Eqs.~\eqref{eqn-eigenvalues-F0}-\eqref{eqn-eigenvalues-F1},
\begin{align}
\frac{e_p}{e_p'} \leq 
\frac{
1
}{
\min\left(D_1,\frac{1}{D_1},\ldots,D_d,\frac{1}{D_d},1\right)
}
\end{align}
where the numerator ($=1$) corresponds to the upper bound on the first ratio.
This proves \myeqnref{eqn-ep-suboptimal1}.

\end{document}